\documentclass{article}

\usepackage[english]{babel}  
\usepackage[top=2cm, bottom=3cm, left=2cm, right=2cm]{geometry}
\usepackage{graphicx}
\usepackage{amsmath}
\usepackage{amssymb}
\usepackage{upgreek}
\usepackage{chemmacros} 
\usepackage{hyperref}
\hypersetup{
  colorlinks=true,
  linkcolor=blue,
  urlcolor=blue,
  filecolor=blue,
  citecolor=blue
}
\usepackage{cleveref}
\usepackage{float}
\usepackage{grffile}
\usepackage{txfonts}
\usepackage{booktabs}
\usepackage{dsfont}
\usepackage{subcaption}
\usepackage{array}
\usepackage{multirow}

\title{Engine position effects on contrail evolution for a realistic aircraft configuration}
\author{}
\date{}

\begin{document}
\maketitle

\noindent {\large Rémy Annunziata$^{\star}$, Nicolas Bonne$^{\star}$, François Garnier $^\ddag$}\\

\noindent \textbf{Abstract:} This study investigates the influence of representative engine positions on contrail evolution during the vortex and dissipation regimes, using three-dimensional simulations of a realistic aircraft geometry. Large Eddy Simulations are employed, coupled with an Eulerian microphysical bulk model, and initialized using fields obtained from prior Reynolds-Averaged Navier-Stokes simulations. This approach enables a consistent transition from near-field jet-vortex interactions to far-field wake dynamics. Three engine placements are examined under two atmospheric stratifications and two relative humidity conditions. The results reveal that engine position influences the onset and evolution of vortex instabilities, alters the descent of the vortex pair, and leads to slight changes in the distribution of particles within the wake. Despite these aerodynamic differences, the microphysical properties of the contrails tend to converge over time for the parameters and configurations covered in this study. From a broader perspective, engine placement strongly influences initial contrail formation and early vortex-regime dynamics. At later stages, these differences largely disappear, as vortex dynamics and atmospheric conditions dominate over the initial dilution changes induced by engine position.\\

\noindent \rule{\linewidth}{.5pt}

\section{Introduction}
\label{sec:section1}

Aircraft-induced condensation trails (contrails) represent a significant factor in aviation’s non-CO\textsubscript{2} environmental impact \cite{Lee09}. Once formed, contrails affect the Earth's radiative balance through dual mechanisms: they reflect incoming solar radiation while simultaneously trapping terrestrial infrared radiation \cite{Burkhardt11}. An intricate interplay between ambient atmospheric conditions and aircraft-specific factors dictates their formation and evolution, resulting in substantial variability in their spatial extent, lifetime, and optical properties. This variability poses challenges in precisely quantifying their net climatic impact. Nevertheless, current assessments suggest that the best estimate effective radiative forcing from contrail cirrus (i.e., persistent contrails) is about 1.67 times that of aviation CO\textsubscript{2}, albeit with substantial uncertainty \cite{Lee21}. Given their substantial contribution to anthropogenic climate forcing, research on contrails focuses on refining our understanding of the mechanisms governing their formation and dissipation and identifying viable mitigation strategies to minimize their climatic footprint \cite{Karcher18}.

The early-stage evolution of a contrail is inherently tied to the dynamics of the aircraft wake. This evolution can be classified into four distinct regimes \cite{Hoshizaki72}. The first of these stages is the jet regime, during which the contrail is formed. While the engine exhaust dilutes with ambient air, gases and particles cool sufficiently to act as nucleation sites. Water vapor, either emitted directly by the engines or already present in the atmosphere, condenses onto these nuclei and freezes under the given thermodynamic conditions to form ice crystals. Potential nucleation sites include soot particles \cite{Karcher96}, generated from kerosene combustion in fuel-rich conditions, as well as potentially on secondary aerosols \cite{Karcher98}. The wake evolution, in this regime, is primarily governed by interactions between the propulsive jet and the wingtip vortex. Its study involves a range of modeling approaches. Box models resolve detailed microphysical processes \cite{Rojo15, Bier22}, but are typically driven offline using prescribed mean dilution rates or trajectories derived from flow simulations. As a consequence, they have limited capability to represent the wake dilution accurately and to capture its feedback on microphysics. By contrast, jet-regime simulations based on Computational Fluid Dynamics (CFD), such as Reynolds-Averaged Navier-Stokes (RANS) \cite{Guignery12, Khou16} or Large-Eddy Simulation (LES) \cite{Paoli13, Ferreira24}, provide a more realistic description of aircraft-induced dilution with online microphysics coupling. However, the representation of the microphysical processes remains simplified in these approaches.

Once the propulsive jet is entirely under the vortex dynamics, the vortex regime begins. A fraction of the ice crystals becomes entrained within the descending vortex pair. The gases undergo adiabatic compression, which heats the particles and causes sublimation. Meanwhile, another portion of the ice crystals ascends back to flight altitude, driven by the upward convection of vorticity and thus forming the secondary wake. This separation is largely influenced by the interaction between the wake and the surrounding atmosphere. Baroclinic instabilities emerge as a result of misalignment between pressure and density gradients, leading to differential vorticity advection that enhances vertical mixing \cite{Spalart96}. As the vortex dipole continues its descent, it becomes susceptible to various instabilities. These include long-wave instabilities, such as the Crow instability \cite{Crow70}, and short-wave instabilities, like the Widnall instability \cite{Widnall74}. The onset of these instabilities signals the transition to the dissipation regime, during which the vortex dipole undergoes progressive breakdown. The vortex and dissipation regimes are typically analyzed using LES \cite{Lewellen96, Chlond98}, while often omitting the simulation of the jet regime. The vortex regime is typically initialized using an analytical Lamb-Oseen-type vortex combined with a Gaussian round jet \cite{Unterstrasser08, Paugam10, Naiman11}. Some studies, however, employ more realistic initial conditions, such as jet-vortex interactions derived from LES \cite{Picot15} or two-dimensional flow fields obtained from realistic aircraft configurations \cite{Huebsch06, Lewellen14pt1}. Finally, the last regime, the dispersion regime, begins once the wake vortices have fully dissipated. In this regime, its behavior is dictated by ambient atmospheric conditions, including wind shear, atmospheric turbulence, and thermodynamic conditions \cite{Unterstrasser10pt2,Lewellen14pt2}.

Contrail evolution during the first three regimes (jet, vortex, and dissipation) is therefore primarily governed by soot number emission index \cite{Karcher09a} and atmospheric conditions such as ambient temperature or humidity \cite{Unterstrasser10pt1,Lewellen14pt1}. These factors are expected to play a particularly important role in determining the number of ice crystals that survive vortex descent, which in turn strongly conditions the radiative impact of the contrail \cite{Unterstrasser10pt2}. Although they play a more secondary role, the characteristics of the aircraft that generate the contrails also influence their evolution \cite{Jessberger13}. These characteristics correspond to design parameters that can, in principle, be modified. Key design factors include the aircraft size \cite{UnterstrasserGorsch14}, which determines vortex circulation, as well as the relative positioning of the propulsive jets and the wingtip vortices \cite{Garnier97}. From this perspective, the interaction between the propulsive jets and the wingtip vortices warrants particular attention, as it can be altered by modifying the engine’s position beneath the aircraft wing. Adjusting this position changes how the plume disperses within the aircraft’s wake, affecting parameters such as species concentration, temperature, and dispersion patterns \cite{Paoli03}. Additionally, the presence of a hot jet has significant aerodynamic effects, bringing the vortices closer together and speeding up their descent \cite{Shirgaonkar07}. Most studies on jet-vortex interactions have primarily focused on contrail formation rather than its long-term evolution. Within this context, the refs. \cite{Paoli13, Ramsay24,Annunziata25} have demonstrated that modifying engine placement influences key contrail properties, such as particle activation and the final ice crystal size at the end of the jet regime. The latter two studies have considered a realistic aircraft configuration. Recent research by Saulgeot et al. \cite{Saulgeot23} further underscores the substantial impact of jet-vortex interactions on contrail evolution during the vortex regime, offering valuable insights into their behavior in the later stages of wake development. The study demonstrates that dynamic modifications induced by engine placement affect the contrails optical properties under different atmospheric stratification conditions. In parallel, a separate study by Naiman et al. \cite{Naiman11} indirectly explores similar effects within a fully three-dimensional framework, comparing two distinct engine positions (corresponding to different aircraft with varying initial circulations) as well as a twin-engine configuration. These simulations, which extend up to twenty minutes, reveal a persistent sensitivity of contrail evolution to aircraft configuration over extended timescales.

Limitations appear in both of these last studies addressing jet-vortex interactions during the vortex and dissipation regimes. The first limitation pertains to the study of Saulgeot et al. \cite{Saulgeot23}, which, although directly investigating the influence of engine position, relies on a two-dimensional framework. This assumption limits its applicability due to the inherently three-dimensional nature of wake dynamics through short- and long-wave instabilities, and the subsequent influence on contrail properties \cite{Unterstrasser14}. Furthermore, contrail ice microphysics is described using a simplified diagnostic framework, in which the vapor and ice phases are inferred from a passive scalar, while crystal growth is not explicitly modeled (i.e., the time-dependent growth of individual ice crystals is not resolved; instead, ice mass and effective radius are diagnosed under a local thermodynamic equilibrium assumption). Another limitation, shared in both studies, lies in their initialization at the onset of the vortex regime. They are based on idealized conditions, using a Lamb-Oseen vortex model combined with a round jet, thereby neglecting the near-field effects associated with realistic aircraft wake initial development. The impact of engine relocation on contrail properties at the end of the jet regime is then not considered. These include variations in dilution, ice crystal size, spatial distribution around the vortex, local supersaturation levels, and temperature fields, as highlighted in ref. \cite{Annunziata25}. This limitation in the initialization is, as previously mentioned, typical of studies addressing contrail evolution during the vortex and dissipation regimes. In most cases, the full aircraft geometry is only taken into account during the initial formation of the contrail in the jet regime \cite{Khou15}. However, a recently developed methodology \cite{Bouhafid24,Bouhafid25} offers a means to bridge the gap between near-field simulations (i.e., RANS simulations), which incorporate realistic aircraft geometry, and the far-field simulations (i.e., LES) of the wake and resulting contrail. This advancement enables the use of a fully realistic aircraft configuration, enabling the resolution of key aerodynamic features, such as tail vortices, which can play a significant role in contrail development by promoting additional vortex instabilities \cite{Fabre02,Bouhafid25}. Furthermore, the method captures the wrapping of the jet around the vortex, a behavior that deviates from the idealized round jet assumption and more accurately represents the actual jet-vortex interactions occurring in the near field. Naturally, this method remains subject to several limitations, mainly from the simplified treatment of the disperse phase through a bulk microphysical representation \cite{Huebsch06,UnterstrasserSolch10}, which does not allow for an accurate representation of the crystal size distribution, and some limitations arising from the intrinsic constraints of the RANS and LES framework. It should be noted that the importance of accounting for such dynamical effects on microphysical properties remains difficult to quantify precisely. However, Garnier et al. \cite{Garnier14} highlight the influence of compressibility effects on crystal size, while Bouhafid et al. \cite{Bouhafid25} demonstrate the importance of including tail vortices in determining plume dimensions.

In this study, we aim to examine how engine position influences contrail evolution during the vortex and dissipation regimes, while consistently accounting for a realistic aircraft geometry and for the effects of engine relocation on contrail formation in the jet regime. To this end, LES are carried out, initialized from preceding RANS computations \cite{Annunziata25}, which explicitly model contrail formation for several engine positions using a fully three-dimensional representation of the aircraft. This approach captures the full aerodynamic complexity of the configuration, including the modified jet-vortex interactions induced by changes in engine location. The LES initialization employs the transition methodology proposed by Bouhafid et al. \cite{Bouhafid24}, which enables a consistent coupling between the RANS simulations in the near-field and the LES in the far-field. In the present study, the analysis is conducted for a single aircraft configuration with three distinct engine positions, under two different atmospheric stratification conditions on contrail development up to five minutes after the end of the jet regime. A primary objective of this work is to assess whether a modification of the initial plume dilution meaningfully alters subsequent contrail development in these later regimes. Particular attention is given to the role of engine positioning in shaping vortex dynamics, plume dispersion, and downstream effects on contrail microphysics. 

The paper is structured as follows: section \ref{sec:meth} details the computational methodology, followed by a description of the numerical configurations in section \ref{sec:numsetup}. The analysis of vortex dynamics and microphysical properties of induced contrails is presented in sections \ref{sec:VxDy}, \ref{sec:Mphi}, and \ref{sec:turb}. Finally, section \ref{sec:ccl} provides conclusions and a discussion of the findings and their limitations.

\section{Methodology}
\label{sec:meth}

\subsection{Flow modeling}

Numerical simulations are performed using CEDRE \cite{Refloch11}, a multi-physics computational platform, and in particular, the CHARME solver, a Navier-Stokes solver specifically designed for compressible and reactive flows. This solver utilizes a cell-centered finite volume method for unstructured meshes. LES are subsequently carried out by solving the filtered compressible Navier-Stokes equations:

\begin{equation}
\frac{\partial}{\partial t} ( \overline{\rho} \tilde{Y}_{k} ) + \frac{\partial}{\partial x_{i}}(\overline{\rho}\tilde{u}_{i}\tilde{Y}_{k}) + \frac{\partial}{\partial x_{i}}\big(\overline{\rho}(\widetilde{u_{i}Y_{k}}-\tilde{u}_i\tilde{Y}_k)\big) = \frac{\partial}{\partial x_{i}} \left (\overline{\rho}D_{k}\frac{\partial \tilde{Y}_{k}}{\partial x_{i}} \right)   + \overline{\dot{\omega}}_{k} \label{eq:eq01}
\end{equation}

\begin{equation}
\frac{\partial}{\partial t} ( \overline{\rho} \tilde{u}_{j} ) + \frac{\partial}{\partial x_{i}}(\overline{\rho} \tilde{u}_{i} \tilde{u}_{j}) + \frac{\partial}{\partial x_{i}}\big(\overline{\rho}(\widetilde{u_{i}u_{j}}-\tilde{u}_i\tilde{u}_j)\big)= -\frac{\partial \overline{p}}{\partial x_{j}} + \frac{\partial \overline{\tau_{ij}}}{\partial x_{i}} + \overline{\rho f_{i}}
\end{equation}

\begin{align}
\begin{split}
\frac{\partial}{\partial t} ( \overline{\rho} \tilde{h} ) + \frac{\partial}{\partial x_{i}}(\overline{\rho}\tilde{u}_{i}\tilde{h}) & = \frac{\partial \overline{p}}{\partial t} + \tilde{u}_i \frac{\partial \overline{p}}{\partial x_i} +  \frac{\partial}{\partial x_{i}}  \Bigg(\overline{\rho}c_{p}\alpha\frac{\partial \overline{T}}{\partial x_{i}}-\overline{\rho} (\widetilde{u_i h}-\tilde{u}_i\tilde{h})\Bigg) \\ & - \frac{\partial}{\partial x_{i}}  \Bigg(\sum_{k}\overline{\rho} \tilde{h}_{k}D_{k}\frac{\partial \tilde{Y}_{k}}{\partial x_{i}} - \overline{\tau}_{ij}\frac{\partial \tilde{u}_i}{\partial x_{j}}\Bigg) + \overline{\rho f_iu_i}
\end{split}
\end{align}

\noindent where $\rho$, $u_i$, $p$, $T$, $Y_k$, $\dot{\omega}_k$, $\tau_{ij}$, and $h$ represent the density, velocity components, static pressure, static temperature, species mass fraction, source term for species $k$, viscous stress tensor, and total enthalpy. The quantities $D_k$, $h_k$, $c_p$, $\alpha$, and $f_i$ denote, respectively, the species diffusion coefficient, species enthalpy, specific heat capacity at constant pressure, thermal diffusivity, and body-force components per unit mass. Dynamic viscosities are computed using Sutherland’s law, from which the kinematic viscosity and the diffusion coefficient are derived through the Schmidt number. The specific heat capacity at constant pressure is evaluated using a temperature-dependent polynomial, while the thermal diffusivity is obtained from Eucken’s law. The corresponding values can be found in refs. \cite{Guignery12,Khou15,Khou16}. The species considered in the simulation are air, represented by a single mass fraction $Y_{air}$, water vapor $Y_{H_2O}$, and ice, represented within the bulk scheme by $Y_{H_2O,s}$.

Spatial filtering is applied to variables denoted as $\overline{\Phi}$ through a convolution operation, while variables expressed as $\tilde{\Phi} = \overline{\rho\Phi} / \overline{\rho}$ are filtered using a Favre average, a formulation particularly suitable for compressible flows. The subgrid-scale (SGS) terms that emerge from the spatial filtering of the Navier-Stokes equations, $\tau^{sgs}_{ij}=\overline{\rho}(\widetilde{u_{i}u_{j}}-\tilde{u}_i\tilde{u}_j)$, represent momentum fluxes due to unresolved scales and are modeled using the Smagorinsky model, based on the Boussinesq hypothesis:

\begin{equation}
    \tau^{sgs}_{ij} = -2\nu_t\overline{S}_{ij} + \frac{1}{3}\delta_{ij}\tau^{sgs}_{kk} 
\end{equation}

\noindent where $\nu_t$ denotes the turbulent eddy viscosity, defined as $\nu_t = (C_s\Delta)^2 \lvert \overline{S}\rvert$, with $C_s=0.1$ and $\Delta$ representing the filter cut-off length. The term $S_{ij}$ corresponds to the shear rate tensor. Species flux, expressed as $J^{sgs}_{i,k}=\overline{\rho}(\widetilde{Y_ku_{i}}-\tilde{Y}_k\tilde{u}_i)$, are modeled following a turbulent Fickian diffusion approach, while a turbulent Fourier law describes subgrid enthalpy fluxes as $q^{sgs}_i=\overline{\rho}c_p(\widetilde{Tu_{i}}-\tilde{T}\tilde{u}_i)$. These fluxes are computed using a gradient assumption:

\begin{equation}
     J^{sgs}_{i,k}=-\frac{\nu_t}{Sc_t}\frac{\partial \tilde{Y}_k}{\partial x_i}
\end{equation}

\begin{equation}
     q^{sgs}_i=-\frac{\nu_t}{Pr_t}\frac{\partial \tilde{T}}{\partial x_i}
\end{equation}

\noindent where $Sc_t$ and $Pr_t$ denote the turbulent Schmidt and Prandtl numbers, respectively, both set to 0.9. Time integration is performed using a second-order implicit Runge-Kutta scheme, while the resulting linear system is solved using the Generalized Minimal Residual (GMRES) method. A second-order numerical scheme is applied for diffusive flows. These choices will be discussed further in the discussion section (Section \ref{sec:ccl}) at the end of the paper.

\subsection{Microphysical modeling}

An Eulerian bulk microphysical framework, initially developed by the refs. \cite{Guignery12, Khou15}, is employed to model the microphysical processes. This approach consists of transporting the zeroth- and third-order moments of the particle size distribution. The zeroth-order moment corresponds to the number density of particles within the flow. Here, the particles are assumed to be soot, which then provides the only available sites where ice is present. This assumption is justified under soot-rich rich-burn engine regimes \cite{Kleine18}. 

Soot particles are considered as passive tracers of the flow, under the asymptotic limit of Stokes numbers that tend to zero. This assumption is justified by the Stokes number being on the order of $\sim10^{-4}$ in studies of the vortex and dissipation regimes \cite{Naiman11}. Accordingly, they are considered to be in dynamical equilibrium, with no explicit drag contribution, and in thermal equilibrium with the carrier phase, such that their temperature is equal to that of the surrounding fluid. Under the LES framework, using a Smagorinsky-type subgrid-scale model and a gradient-diffusion hypothesis, the transport equation for the filtered soot particle number density $\tilde{N}_p$ reads:

\begin{equation}
\frac{\partial}{\partial t} ( \overline{\rho} \tilde{N}_{p} ) + \frac{\partial}{\partial x_{i}}(\overline{\rho}\tilde{u}_{i}\tilde{N}_{p}) = \frac{\partial}{\partial x_{i}} \left (\overline{\rho}\Big(D_s+\frac{\nu_t}{Sc_t}\Big)\frac{\partial \tilde{N}_{p}}{\partial x_{i}} \right)
\end{equation}

\noindent where $D_s$ represents the molecular diffusion coefficient of soot particles, associated with the Schmidt number $Sc$, which is prescribed as unity with respect to the kinematic viscosity of air, consistent with the passive tracer hypothesis. In this approach, particles are assumed to be monodisperse within each mesh cell. The smallest admissible radius is thus the radius of an individual soot particle, $r_s$. 

Ice (denoted $H_2O,s$) is represented by the third moment of the particle size distribution. Therefore, it is the ice water content (IWC), $IWC=\rho Y_{H_2O,s}$, that is transported by the flow, where $Y_{H_2O,s}$ is the ice mass fraction. Ice production or loss is governed by phase changes (deposition and sublimation) through the reversible reaction between atmospheric water vapor $Y_{H_2O,v}$ and ice $Y_{H_2O,s}$: $H_2O,v \leftrightarrow H_2O,s$. Consequently, ice production is accounted for by a source term $\overline{\dot{\omega}}_{H_2O,s}$ in the IWC transport equation:

\begin{equation}
    \overline{\dot{\omega}}_{H_2O,s}=\frac{4\pi \tilde{N}_p r_p D_{H_2O,v}M_{H_2O}}{R\overline{T}}\big(\overline{p}_{H_2O,v}-p_{H_2O}^{sat,s}(r_p,\overline{T})\big)G(r_p)\Pi(p_{H_2O}^{sat,l}(\overline{T}),r_p) \label{eq:eq08}
\end{equation}

\noindent where $r_p$ denotes the ice crystal radius. Ice crystals are therefore assumed to be monodisperse within each cell, just like soot. Note that $N_p$ soot particles are conserved in the calculation even when the surrounding ice has sublimated. Thus, it is the presence or absence of ice within a cell that determines whether the particles are ice crystals or just soot. 

The ice crystal radius value is determined from the mass fraction of ice $Y_{H_2O,s}$, by applying fundamental geometric considerations under the assumption that ice crystals adopt a spherical morphology. The corresponding expression is given by:

\begin{equation}
    r_p=\bigg(r_s^3+\frac{3}{4\pi}\frac{\overline{\rho} \tilde{Y}_{H_2O,s}}{\overline{\rho}^p_{H_2O,s}(\overline{T})\tilde{N}_p}\bigg)^{1/3}
\end{equation}

\noindent where $\overline{\rho}^p_{H_2O,s}(\overline{T})$ represents the mass density of ice within a particle, taken from ref. \cite{Tabazadeh00}. 

From Equation \ref{eq:eq08}, $D_{H_2O,v}$ denotes the diffusion coefficient of water vapor, which is given by ref. \cite{Hall76}. The terms $\overline{p}_{H_2O,v}$, $p_{H_2O}^{sat,s}(\overline{T})$ and $p_{H_2O}^{sat,l}(\overline{T})$ correspond, respectively, to the partial pressure of water vapor, the saturation pressure with respect to ice, and the saturation pressure with respect to liquid water \cite{Murphy05}. The Kelvin effect is also incorporated for calculating saturation around the particles.

For the last terms in Equation \ref{eq:eq08}, the function $G(r_p)$ accounts for the transition between the kinetic gas regime and the diffusion regime \cite{Davies76}. The function $\Pi(p_{H_2O,v}^{sat,l},r_p)$, given by:

\begin{equation}
    \Pi(p_{H_2O}^{sat,l},r_p) = 
    \begin{cases}
        0 & \text{if}~~ p_{H_2O,v} \le p_{H_2O}^{sat,l} ~~\text{and}~~ r_p = r_s \\
        1 & \text{if}~~ p_{H_2O,v} \ge p_{H_2O}^{sat,l} ~~\text{or}~~ r_p > r_s
    \end{cases}
\end{equation}

\noindent governs the deposition on particles.

It should be noted that both this formulation and the monodisperse closure adopted here exhibit intrinsic limitations for representing the disperse phase. As shown in the ref. \cite{Huebsch06,UnterstrasserSolch10}, although the bulk formulation provides a reasonably accurate estimate of the ice mass within a contrail, it tends to substantially underestimate the number of ice crystals surviving the vortex regime. This limitation is even more pronounced when a monodisperse closure is employed, as such an approach suppresses competition through the Kelvin effect. These limitations should therefore be kept in mind when interpreting the present results. The implication of microphysical modeling choices and assumptions on the results will be further discussed in Section \ref{sec:ccl} at the end of the article.

\subsection{Transition from near-field to far-field calculation}

To conduct a comprehensive assessment of engine installation effects on a realistic aircraft configuration, from contrail formation through its subsequent evolution up to the end of the dissipation regime, it is interesting to simulate the initial formation phase. This approach makes it possible to directly simulate the resulting changes in dilution and their impact on microphysical properties, without requiring assumptions regarding these properties (microphysical or dynamical) at the onset of the vortex regime, and to ensure a consistent coupling between the near-field and far-field dynamics. Especially given that ref. \cite{Annunziata25} have shown that engine position can modify the ice crystal radius, potentially leading to differences in the number of crystals surviving the vortex regime. Moreover, \cite{Bouhafid25} have demonstrated that plume spreading is also significantly affected by the horizontal tailplane vortices. Such a task presents significant challenges, primarily due to the distinct spatial and temporal scales characterizing each regime and the complexity of the underlying physical processes \cite{Paoli16}. To address these challenges, a novel methodology has been recently developed and validated by Bouhafid et al. \cite{Bouhafid24}. This approach effectively captures both the near-field and far-field dynamics of contrail evolution. In the near field, a RANS framework is employed, inspired by the work of Khou et al. \cite{Khou15}, which is particularly well-suited for representing the aircraft geometry and thus ensuring a realistic description of the initial dilution process. Meanwhile, the far-field dynamics, encompassing vortex and dissipation regimes, are modeled using LES to resolve unsteady phenomena such as vortex instabilities. This hybrid methodology seamlessly integrates the RANS and LES approaches, effectively bridging the near-field and far-field regimes.

In the initial phase of this methodology, 3D RANS simulations are performed over a computational domain extending approximately twenty wingspans downstream of the aircraft. Subsequently, a two-dimensional slice of the RANS solution is extracted, oriented perpendicular to the flight direction. This extraction occurs at a sufficient distance from the aircraft, ensuring that axial gradients remain negligible in comparison to transverse gradients, a necessary condition for applying periodic boundary conditions \cite{Kolomenskiy18}. The flight speed is then subtracted from the velocity field, which is thus expressed in the ground reference frame. A new computational mesh is then generated, onto which the RANS solution is interpolated. The extracted slice is then extruded over a length corresponding to the Crow instability wavelength, given by $\lambda_{crow}=8.6b_0$ \cite{Crow70}, where $b_0$ represents the spacing between the counter-rotating vortices. This extrusion provides a sufficiently representative domain for capturing and analyzing the temporal development \cite{Lewellen01}.

To ensure a seamless transition between the turbulence characteristics in the RANS field and the resolved turbulence in the LES framework, synthetic fluctuations are introduced using the Modified Discretizing and Synthesizing Random Flow Generation (MDSRFG) method \cite{Castro13}. These fluctuations are superimposed onto the jet to enhance its turbulent structure, ensuring consistency in turbulence representation across the modeling approaches. Additionally, another set of fluctuations is introduced to account for atmospheric turbulence. These fluctuations are generated randomly from jet and atmospheric turbulence spectra, using the turbulence magnitude and the RANS velocity field as input. These perturbations play a crucial role in contrail evolution, as they contribute to the destabilization of vortical structures and are therefore intrinsically linked to their eventual dissipation \cite{Picot15}. 

A further essential aspect of the initialization process is the inclusion of gravitational effects to enforce atmospheric stratification. While gravity and stratification are typically neglected in the near-field due to their minimal impact on the early plume dynamics, stratification becomes increasingly important in the far-field regime. Its inclusion leads to variations in pressure, temperature, and consequently, relative humidity with altitude, driven by a specified Brunt-Väisälä frequency. This stratification effect has been shown to play a critical role in contrail evolution, as demonstrated in the studies by the refs. \cite{Unterstrasser08, Naiman11}.

\section{Numerical Setup}
\label{sec:numsetup}

As previously mentioned, the present study focuses on the vortex and dissipation regimes. The jet regime, in contrast, has been extensively analyzed in a previous study that examined the impact of near-field engine positioning on contrail formation. Consequently, the computational results from the RANS simulations of Annunziata et al. \cite{Annunziata25}, in which soot particles are assumed to be fully activated, will serve as the initial conditions for the LES conducted in this study. The results obtained with this modeling approach are retained instead of those based on soot activation by adsorption, since, as discussed in \cite{Annunziata25}, they are in much better agreement with available measurements and with previous modeling studies. It is furthermore expected that activation processes are primarily relevant in the near field, while exerting a much weaker influence farther downstream, particularly on crystal sublimation during the vortex regime. The use of these RANS fields is therefore important, as they enable the inclusion of a fully realistic aircraft configuration and capture the relevant near-field phenomena, within the scope and limitations of RANS modeling. While the latter enables robust simulations of contrail formation by incorporating realistic aircraft geometry and accurately capturing jet-vortex interactions, it also entails inherent limitations, including the inability to represent unsteady phenomena and the near-complete modeling of turbulent structures.

\begin{figure}[!ht]
\centering
\includegraphics[width=0.8\linewidth]{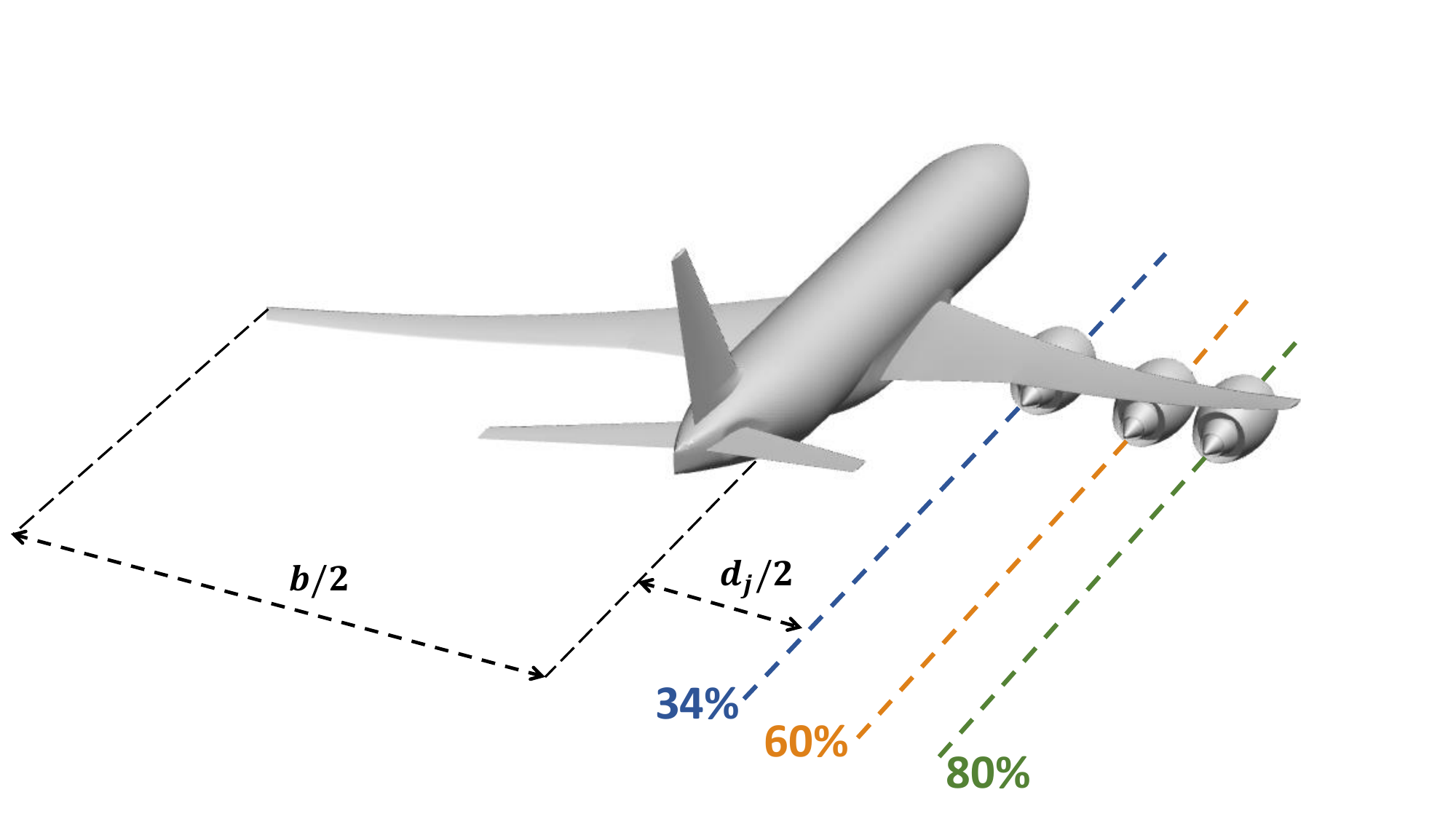}     
\vspace{-0.2cm}
\caption{CRM engine configurations: $d_j / b=$ $0.34$ (blue); $0.60$ (orange); $0.80$ (green).}
\label{fig:01}
\end{figure}

The aircraft configuration used in the RANS simulation is the Common Research Model (CRM), a geometry developed by NASA, which is representative of a Boeing 777 (Figure \ref{fig:01}). This complete configuration includes the fuselage, wings, engines, and tailplane. The pylon connecting the wing and the engine has been removed to facilitate engine relocation. The aircraft features a wingspan of $b=58.8\text{ m}$ and an average chord length of $c_m=6.5$ m. Three distinct engine placements are analyzed in the previous study, and the present work extends this investigation. The first configuration corresponds to the original CRM engine position, with a relative spanwise placement of $d_j/b = 0.34$, meaning that the engines are positioned at 34\% of the half-wingspan. This configuration closely resembles engine placements found on aircraft such as the Airbus A320, A350, or Boeing 777. The second configuration places the engines at $d_j/b = 0.60$, corresponding to 60\% of the half-wingspan, representative of the outboard engine position on a B-747. The third configuration, at $d_j/b = 0.80$, aligns the jet exhaust with the wingtip vortex. This placement corresponds to the case where the two wingtip vortices, which under elliptic lift distribution are separated by a distance $b_0 = \frac{\pi}{4}b$ \cite{Jacquin01}, positioning the jet at approximately 80\% of the wing’s half-span. This engine position has also been shown to influence the optical properties of the contrail, leading to reduced extinction compared to alternative positions during the vortex phase \cite{Saulgeot23}. For brevity, the thermodynamic conditions of the engine and the RANS computational domain are not detailed in this paper. Readers interested in these specifics are encouraged to refer to the near-field study for further information \cite{Annunziata25}. As shown in Figure \ref{fig:01} and for the sake of clarity throughout this article, the 34\% configuration is consistently represented in blue, the 60\% configuration in orange, and the 80\% configuration in green.

\begin{figure}[H]
\centering
\includegraphics[width=1.\linewidth]{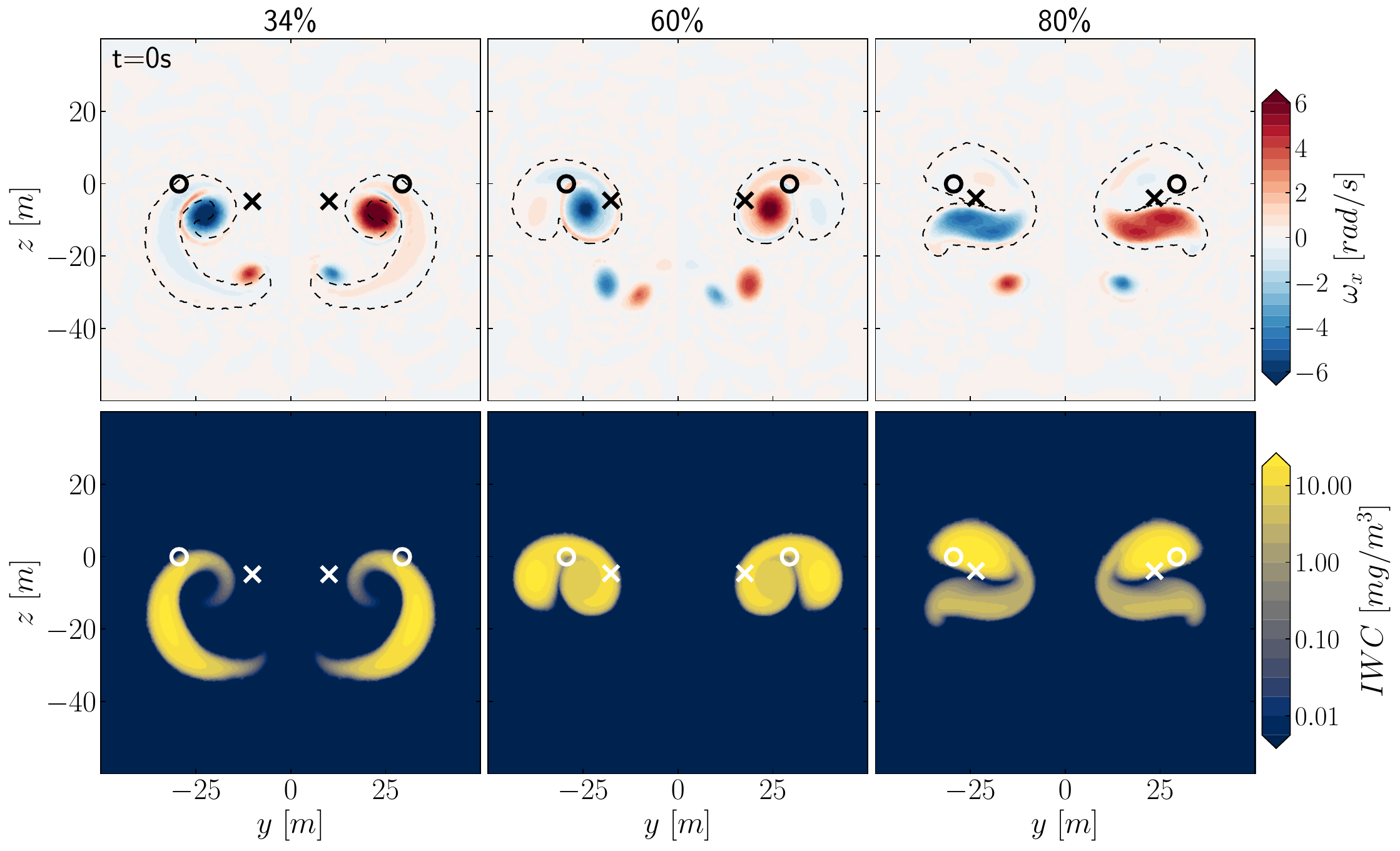}
\caption{RANS slices at ${x/b} {=} {18}$ for LES initialization: streamwise vorticity ${\omega}_{{x}}$ and ${IWC}$, with initial engine (cross) and vortices positions (circle), for ${d_j/b}$ = 0.34 (left), 0.60 (center), 0.80 (right).}
\label{fig:fig02}
\end{figure}

Following the recommendations of Bouhafid et al. \cite{Bouhafid24}, the two-dimensional slice is extracted at $x/b=18$ for each configuration, corresponding to a position 18 wingspans downstream of the engine exit. The sensitivity analysis conducted in the study indicates minimal variation in results based on the slice extraction location, as demonstrated by their comparative assessment between $x/b=15$ and $x/b=18$. This distance corresponds to the location at which the propulsive jet velocity becomes negligible compared with the transverse vortex-induced velocities (see Fig. 5 in \cite{Annunziata25}). The slices extracted at $x/b=18$ for each configuration are shown in Figure \ref{fig:fig02}, where the streamwise vorticity field $\omega_x$ is displayed along with the IWC isoline at $10^{-3}\text{ mg}/\text{m}^3$. The lower row also presents the IWC field. The initial positions of the engine and the wingtip, where the vortex originates, are indicated on both fields.

In all configurations, the wingtip vortex remains the dominant wake feature, corresponding to the peak vorticity. In the 34\% engine position case, the two additional vortices correspond to the tail vortices and exhibit lower vorticity levels than the wingtip vortex. As a reminder, and as visible in the slices, modifying the engine position in the CRM configuration induces flow separation over the wing, which in turn leads to the formation of additional vortex structures in the 60\% and 80\% cases. In the 60\% configuration, the tail vortices have reached their lowest vertical position, and the two remaining vortices visible in the slice then correspond to the lift-off vortices. Finally, in the 80\% configuration, the merging vortices can be observed along with the tail vortices positioned below. For each configuration, the corresponding vortex spacing $b_0$ is as follows: $b_{0,34\%}=46$ m; $b_{0,60\%}=47$ m and $b_{0,80\%}=46$ m. In the latter case, since the slice at $x/b=18$ corresponds to the vortex merging stage, the value of $b_0$ is thus defined as the barycentric distance between the centers of the two merging vortices.

A particularly noteworthy aspect of these initializations is the distinct morphology of the jets in each case. The jets exhibit significant variations in shape, wrapping around the vortices in different configurations. This phenomenon has been extensively analyzed in the ref. \cite{Annunziata25}, and it is intrinsically linked to near-field aerodynamics. It is interesting to note that at the initialization point, the distance between the vortex core and the jet center is 14 m for the 34\% configuration, 13 m for the 60\%, and 12 m for the 80\%. These values differ from the initialization distances used by Saulgeot et al. \cite{Saulgeot23}, as their analysis does not account for the near-field interactions. 

Regarding the microphysical properties of the plume at initialization, as in the RANS computations, all soot particles are converted into ice crystals, the emission index of the ice crystals $\text{EI}_{\text{ice}}$ is equal to that of soot, $1\times10^{15} \text{ kg-fuel}^{-1}$. These soot particles have a radius of 27 nm, and for each configuration, the mean radii $r_{p,0}$ and total mass per meter of flight at initialization $m_{i,0}$ are listed in Table \ref{tab:tab01}, which summarizes all relevant parameters for the LES simulations.

\begin{table}[H]
    \centering
    \caption{LES simulation parameters}
    \label{tab:tab01}
    \begin{tabular}{l c c}
        \hline
        \textbf{Category} & \textbf{Parameter} & \textbf{Value} \\
        \hline
        \multicolumn{3}{c}{\textbf{Mesh resolution}} \\
        \hline
        Axial domain length & $L_x$ & 400 m \\
        Transverse/vertical domain size & $L_y, L_z$ & 1.2 km \\
        Axial grid spacing & $\Delta x$ & 2 m \\
        Transverse/vertical grid spacing & $\Delta_{min} y,~\Delta_{min} z$ & 0.5 m \\
        Grid points & $N_x\times N_y\times N_z$ & 200 $\times$ 552 $\times$ 1429  \\
        \hline
        \multicolumn{3}{c}{\textbf{Jet and vortex properties (34\%; 60\%; 80\%)}} \\
        \hline
        Vortex radius & $r_{c}$ & 5.00; 5.20; 5.20 m \\
        Crow instability wavelength & $\lambda_{crow}$ & 400 m \\
        Jet dissipation rate & $\epsilon_{j}$ & 30; 29; 35 $\text{m}^2/\text{s}^3$ \\
        Jet radius & $r_{j}$ & 4.10; 4.10; 3.50 m \\
        \hline
        \multicolumn{3}{c}{\textbf{Ice crystal properties (34\%; 60\%; 80\%)}} \\
        \hline
        Ice crystal emission index & $\text{EI}_{\text{ice}}$ & $1\times10^{15} \text{ kg.fuel}^{-1}$ \\
        Soot radius & $r_s$ & $27 \text{ nm}$ \\
        Ice mass & $m_{i,0}$ & 0.032; 0.042; 0.037 $\text{ g}/\text{m}$  \\
        Mean radius & $r_{p,0}$ & 842; 920; 884 nm \\
        \hline
        \multicolumn{3}{c}{\textbf{Atmospheric conditions}} \\
        \hline
        Atmospheric dissipation rate & $\epsilon_{atm}$ & $6\times10^{-5} \text{ m}^2/\text{s}^3$ \\
        Relative humidity (ice) & $RH_i$ & 110\%; 120\% \\
        Temperature & $T$ & 223 K \\
        Pressure & $p$ & 264 hPa \\
        Stratification & $N_{BV}$ & 0.012; 0.030 $s^{-1}$ \\
        \hline
    \end{tabular}
\end{table}

\begin{figure}[H]
\centering
\includegraphics[width=0.75\linewidth]{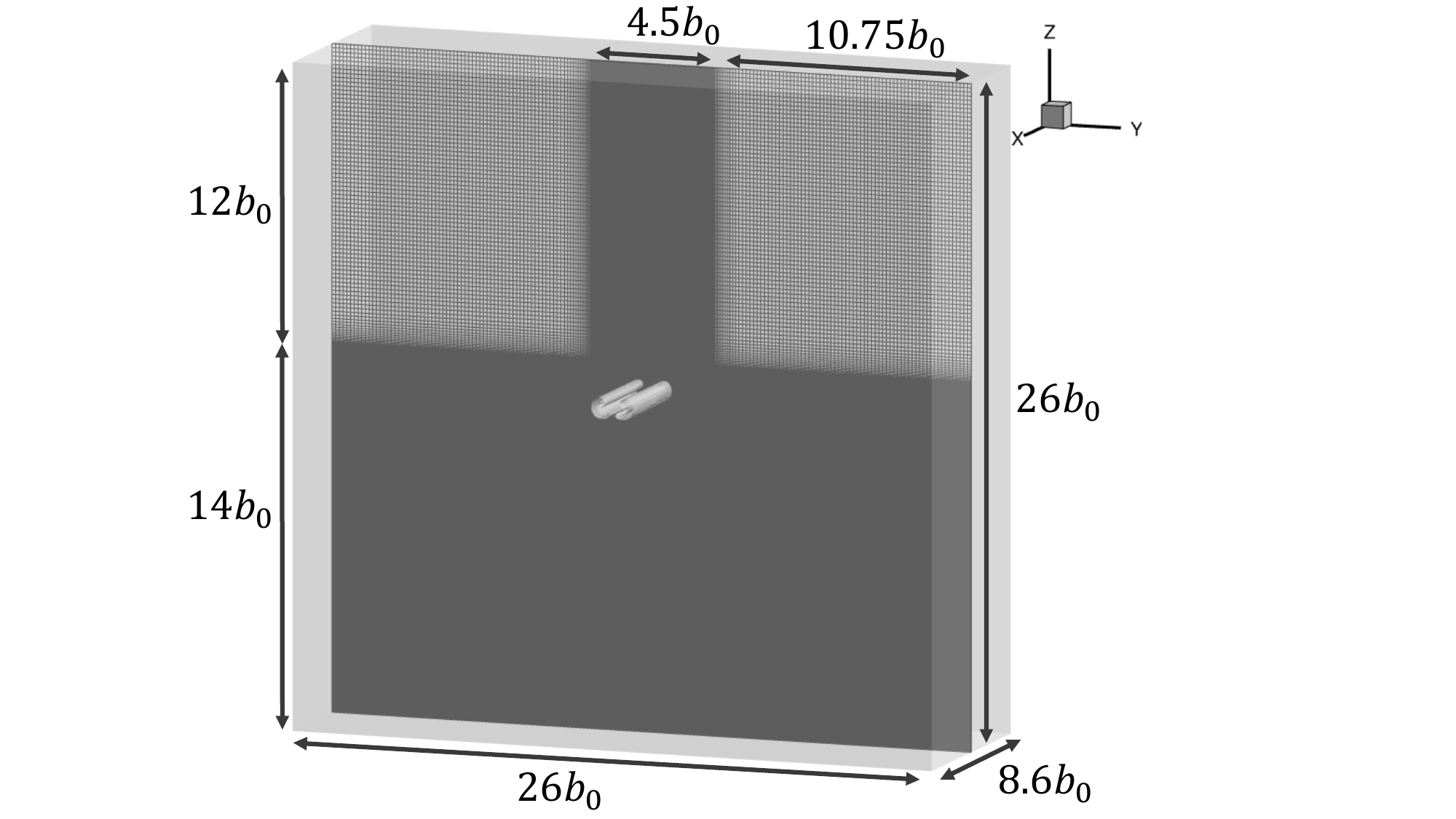}
\vspace{0.2cm}
\caption{LES domain with mesh, and their relevant dimensions.}
\label{fig:fig03}
\end{figure}

The extracted slices are subsequently extruded over a length corresponding to the Crow instability wavelength $\lambda_{crow}=8.6b_0$. Given the close values of $b_0$ across all cases, we assume a single Crow wavelength for the three configurations, expressed as $\lambda_{crow}=8.6\,b_{0,34\%}\approx400$ m, following the refs. \cite{Bouhafid24,Bouhafid25} for the CRM configuration. For the transverse and vertical directions, we set $L_y=L_z=26\,b_0\approx1.2$ km, ensuring minimal influence of the edge boundary conditions on the wake. Regarding the mesh resolution, a grid spacing in the axial direction of $\Delta x=2$ m is used. This results in 200 points along the axial direction, which is sufficient to resolve the Crow wavelength, and provides 18 points to accurately capture the smallest wavelengths in the range $0.79<\lambda/b_0<1.57$ \cite{Fabre02}. For the discretizations in the transverse and vertical directions, special refinement is applied in the wake vortex descent zone as in ref. \cite{Picot15}, ensuring that the mesh resolution satisfies $\frac{\Delta}{r_c}=0.1$, meaning that 10 grid points resolve each vortex radius $r_c$. As demonstrated in the study by Annunziata et al. \cite{Annunziata25}, engine displacement has a negligible impact on the wingtip vortex radius, particularly at $x/b=18$, where the radii remain very similar across configurations. Therefore, we adopt the minimum vortex radius, $r_{c,34\%}=5$ m, leading to a uniform mesh resolution of $\Delta_{min}=\Delta_{min} y=\Delta_{min} z=0.5$ m in the plume descent region. This region extends 640 m in height and 100 m in width. The mesh is then gradually coarsened away from this core zone, with the outermost cells (located near the domain boundaries) reaching a resolution of $\Delta_{max} y=6.8$ m and $\Delta_{max} z=7.4$ m, thereby optimizing computational efficiency. This meshing strategy enables the use of a single, consistent computational grid across all three engine configurations. The domain and mesh used, with their dimensions, are shown in Figure \ref{fig:fig03}, with the plume of the 34\% case in the center. Periodic boundary conditions are applied in the axial direction, while the remaining boundaries (bottom, top, left, and right) are prescribed with fixed temperature and ambient pressure, along with zero velocity.

As mentioned in the previous section, turbulence properties in the jet are incorporated at LES initialization using values derived from the RANS simulations for each configuration. Since each case exhibits distinct turbulence dissipation rates and jet radii, the corresponding values at $x/b=18$ are as follows: for 34\% $\epsilon_j=30\text{ m}^2/\text{s}^3$,$~r_j=4.1$ m; for 60\% $\epsilon_j=29\text{ m}^2/\text{s}^3$,$~r_j=4.1$ m and for 80\% $\epsilon_j=35\text{ m}^2/\text{s}^3$,$~r_j=3.5$ m.

For the atmospheric conditions in the LES domain, a single turbulent dissipation rate value is used across all simulations, set at $\epsilon_{atm}=6\times10^{-5}\text{ m}^2/\text{s}^3$. As determined from the RANS calculations, the atmospheric flight conditions remain unchanged, with a relative humidity with respect to ice of $RH_i=110\%$, and temperature and pressure values of $223$ K and $264$ hPa, respectively. In addition to this baseline ambient relative humidity, the sensitivity of the results to engine position is also assessed against variations in ambient relative humidity by considering additional cases at 120\%. The following equations then describe the temperature and pressure profiles as a function of altitude $z$ :

\begin{equation}
    \frac{1}{\Theta}\frac{d\Theta}{dz}=\frac{N_{BV}^2}{g}, \quad \frac{dp}{dz}=-g\rho(z)
\end{equation}

\noindent where $\Theta=T\big(\frac{p}{p_0}\big)^{-(\gamma-1)/\gamma}$ represents the potential temperature of the atmosphere, with $p_0$ representing the reference pressure and $\gamma$ denoting the adiabatic index. $g$ is the gravitational acceleration, and $N_{BV}$ is the Brunt-Väisälä (BV) frequency. In this study, two stratification regimes are considered: a moderate stratification with $N_{BV}=0.012~\text{s}^{-1}$ and a strong stratification at $N_{BV}=0.030~\text{s}^{-1}$. These values have been carefully selected to represent realistic atmospheric conditions \cite{Schumann17} while also being relevant based on the findings of Saulgeot et al. \cite{Saulgeot23}. Considering the inverse Froude number defined as $Fr^{-1}=\tau_bN_{BV}$, where $\tau_b=2\pi b_{0}^2/\Gamma$ is the characteristic time of the vortex dipole, the selected values correspond by analogy with their study to values near $Fr^{-1}\approx0.3$ and $Fr^{-1}\approx0.8$. Then, for a given vortex characteristic time, weak and moderate stratification lead to minimal differences between engine positions, whereas strong stratification results in more significant variations in optical impact \cite{Saulgeot23}. 

\begin{figure}[H]
\centering
\includegraphics[width=1.\linewidth, trim=0cm 3.6cm 0cm 2.7cm, clip]{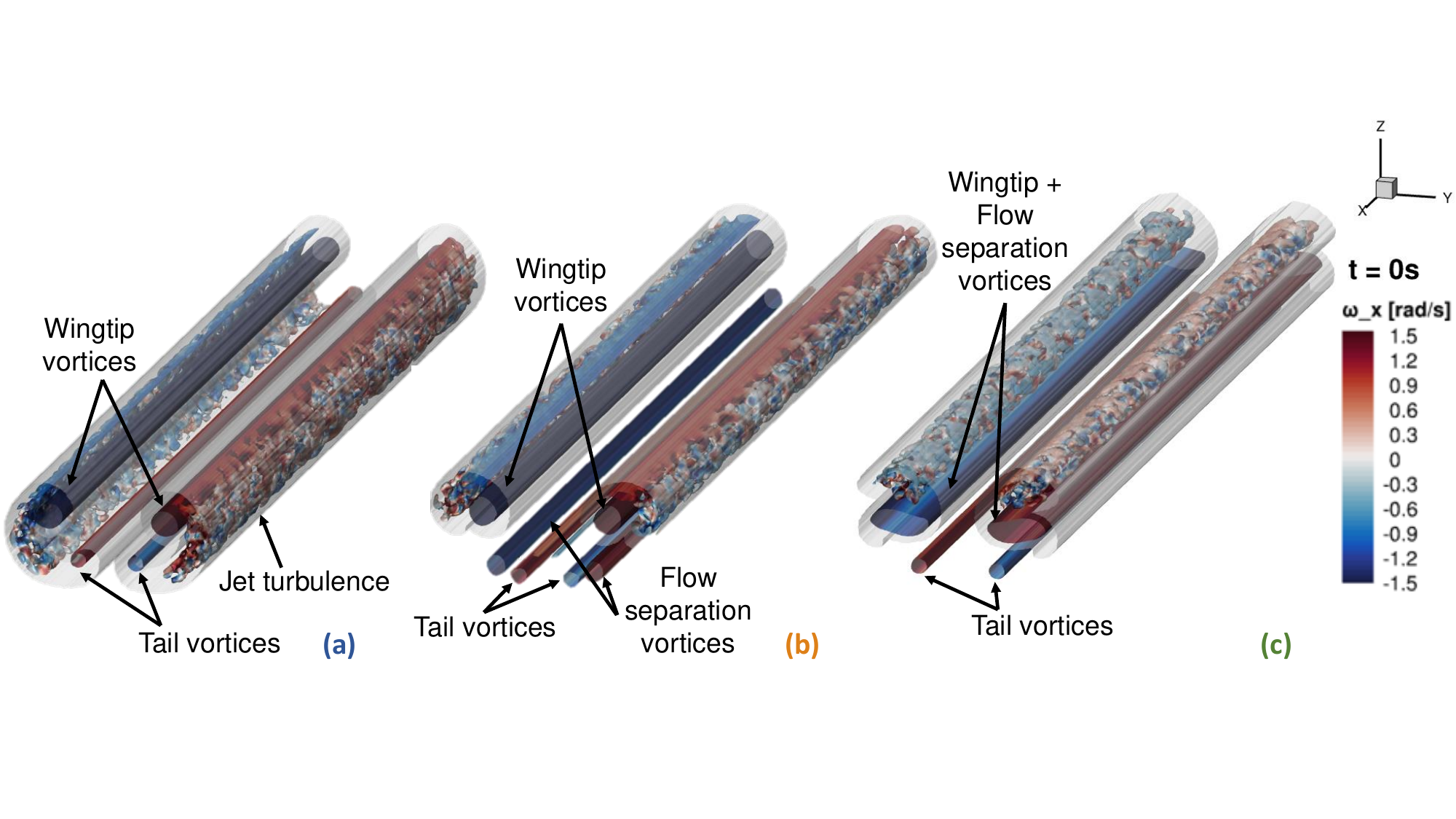}     
\caption{LES initialization: ${Q}$-criterion iso-surface (${Q =} \text{0.03}~\text{s}^{\text{-2}}$) colored by streamwise vorticity, and white IWC iso-surface ($\text{10}^{\text{-3}}~\text{mg}/\text{m}^{\text{3}}$) for ${d_j/b}$ = 0.34, 0.60, 0.80.}
\label{fig:fig04}
\end{figure}

The LES initializations for each configuration are finally illustrated in Figure \ref{fig:fig04}, where the $Q$-criterion iso-surface is plotted at $0.030\text{ s}^{-2}$. This visualization captures the wingtip vortices, tail vortices, wake turbulence, and flow separation-induced vortices. The iso-surface is colored according to the $x$-component of the rotational vector, providing insight into the rotational structures present in the wake. Additionally, the $IWC$ iso-surface at $10^{-3}\text{ mg}/\text{m}^3$ is plotted, offering a qualitative representation of contrail morphology at the onset of the LES computations.

\section{Wake Flows Characteristics}
\label{sec:VxDy}

By modifying the engine's position beneath the aircraft wing, not only the jet properties but also the kinematics of the wingtip vortices are affected. In the near-field, the proximity of the jet to the vortex reduces the amplitude of tangential velocities, thereby modifying the vortex circulation \cite{Ghimire17, Boelle23}. In addition, secondary vortex structures further alter circulation \cite{Annunziata25}. This circulation, together with the vortex spacing $b_0$ (which remains nearly similar across all three configurations), governs the characteristic time of the vortex dipole $\tau_b$. It then varies with the configuration, leading to the following vortex timescales: $\tau_{b,34\%}= 27\text{ s}$; $\tau_{b,60\%}=30\text{ s}$ and $\tau_{b,80\%}=24\text{ s}$. 

\begin{figure}[H]
\centering
\includegraphics[width=1.0\linewidth, trim=0.2cm 4.6cm 0.0cm 4.5cm, clip]{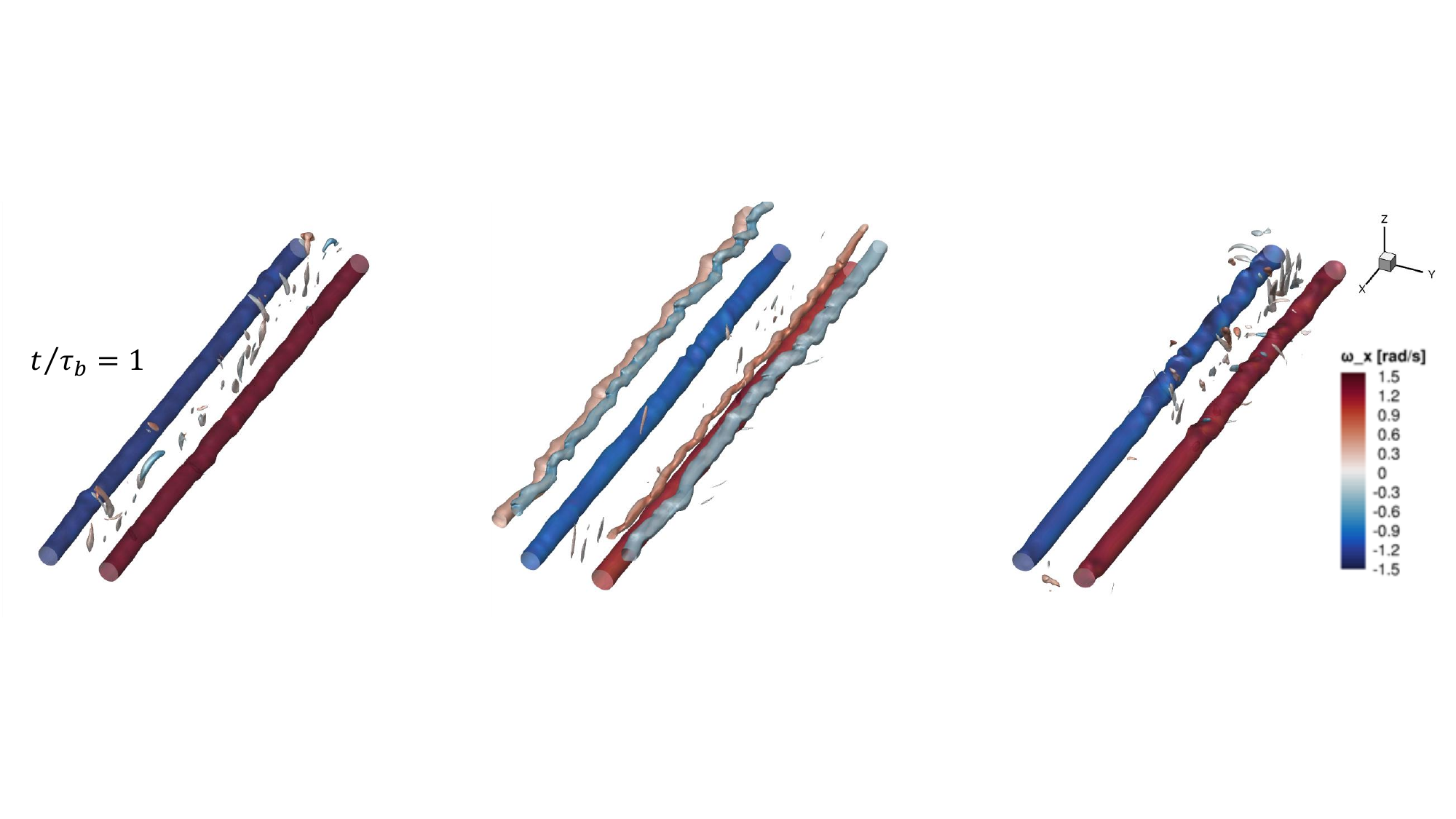}
\includegraphics[width=1.0\linewidth, trim=0.2cm 4.6cm 0.0cm 4.8cm, clip]{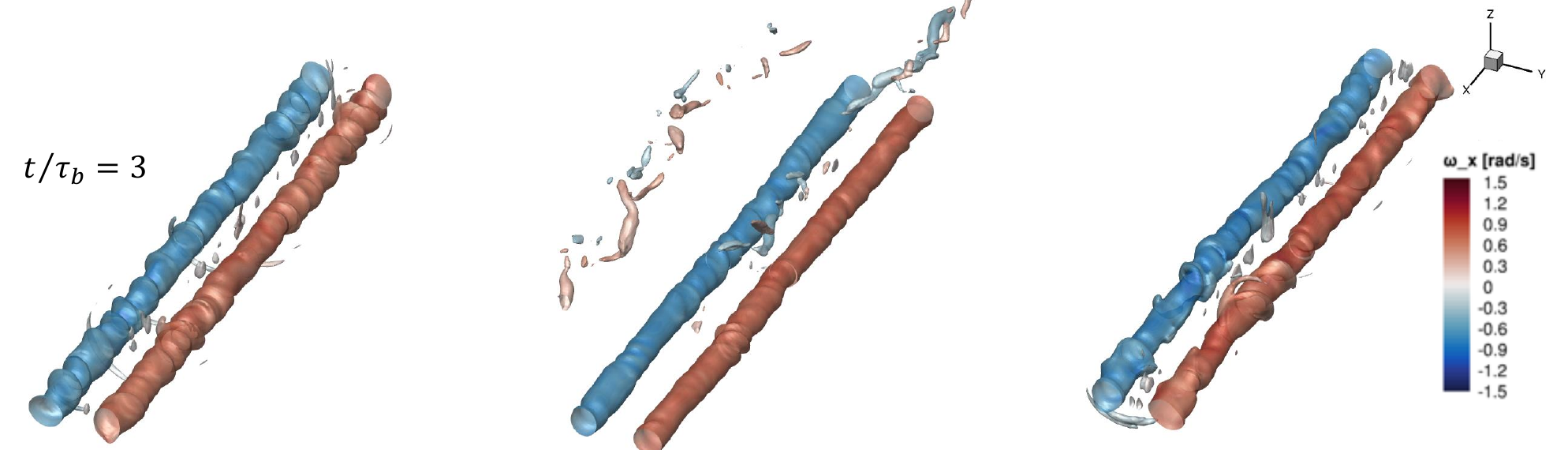}
\includegraphics[width=1.0\linewidth, trim=0.2cm 4.6cm 0.0cm 4.8cm, clip]{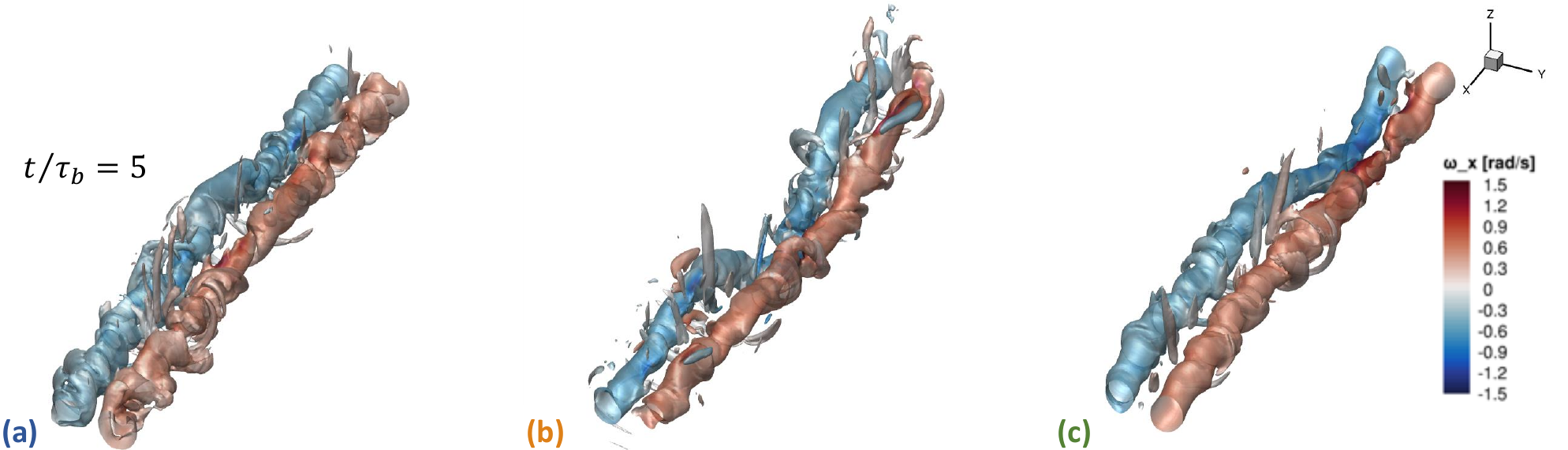}
\caption{${Q}$-criterion iso-surface (${Q=0.03~\text{s}^{-2}}$) evolution, colored by $\omega_x$, for ${N_{BV}=0.012\text{ s}^{-1}}$, at ${t/\tau_b=1,3}$, and ${5}$: ${d_j / b=}$ 0.34 (a); 0.60 (b); 0.80 (c).}
\label{fig:fig05}
\end{figure}

Differences obtained from the RANS calculations, which are used to initialize the LES, particularly in terms of vortex and jet properties, lead to aerodynamic behaviors specific to each engine configuration. The evolution of the vortex dipole for $N_{BV}=0.012\text{ s}^{-1}$ (i.e. $Fr^{-1}\approx0.3$), at three instants, normalized by the characteristic time of the vortex dipole, $t/\tau_b=1$, $3$ and $5$ is shown in Figure \ref{fig:fig05}. The figure displays the iso-surface of the $Q$-criterion at $Q=0.03~\text{s}^{-2}$, colored by the rotational component along the $x$-axis. These visualizations reveal the development of short-wavelength instabilities, starting at $t/\tau_b=1$, and the emergence of the Crow instability at $t/\tau_b=5$ in all three cases, approaching vortex dipole breakdown. These instabilities appear and evolve differently depending on the engine position. Given the rather different initial conditions inherited from the RANS field, many phenomena could account for these differences. 

In the case of short-wavelength instabilities (i.e., elliptic instabilities), bending and double-helix modes are observed in all configurations \cite{ROY11}, the latter being characteristic of flows with axial motion. These modes are further amplified by jet-induced turbulence and axial components, the proximity of which varies by configuration. Such instabilities are particularly prominent in the 60\% and 80\% cases, where their development is already apparent at $t/\tau_b=1$, whereas they are much less marked for the 34\% configuration. However, for the latter and the 80\% case, the tail vortex also plays a key role by interacting with the stress field around the vortices, thereby enhancing instability development \cite{Fabre02}. For the 60\% configuration, the tail vortex interacts with the vortex generated by flow separation and remains near cruise altitude, inducing fewer instabilities in the wake vortices at $t/\tau_b=3$ than the other cases. In 60\%, the vortices originating from the tailplane and those caused by flow separation interact, leading to the development of elliptical instabilities in both structures. This interaction gives rise to the formation of $\Omega$-loops \cite{Ortega03} and their breakdown.

Finally, Crow instabilities develop in all configurations, with different inflection points and growth rates. These are influenced by the interaction between elliptic and Crow instabilities, which can modify the onset and amplification of the latter \cite{Leweke98}. The onset of Crow instabilities is also known to be highly sensitive to the nature of the initial perturbations, as shown by Paugam et al. \cite{Paugam10}. As such, the presence of jets with varying shapes and intensities can initiate and modulate the development of Crow instabilities in different ways. Although not shown here, the vortex evolution under strong stratification $N_{BV}=0.030\text{ s}^{-1}$ (i.e. $Fr^{-1}\approx0.8$) follows a similar instability progression, with initial short-wavelength development followed by the Crow instability, both phenomena occurring more rapidly and intensely due to the enhanced effects of stratification \cite{Holzapfel01}.

\begin{figure}[H]
\centering
\includegraphics[width=1.\linewidth]{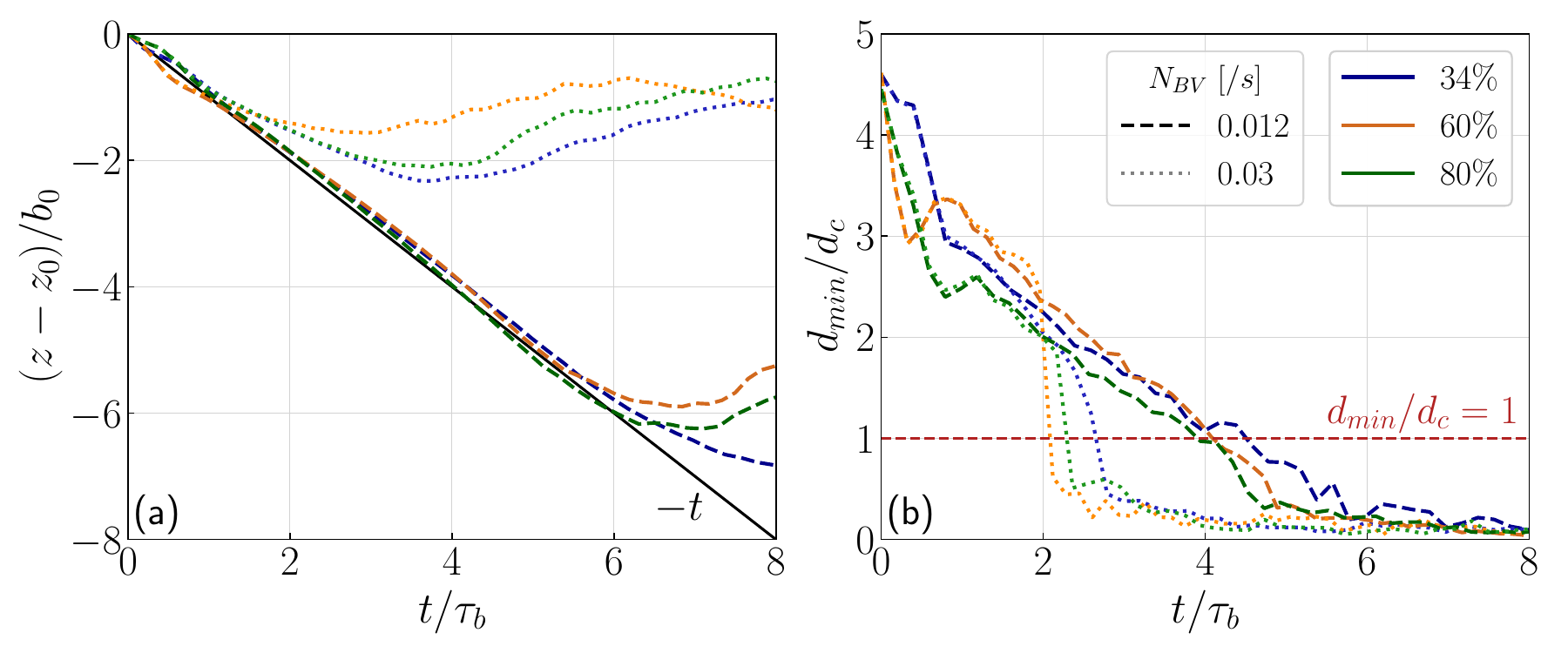}     
\caption{Descent of vortex dipole ${(z-z_0)/}$ (a) and minimum core separation ${d_{\min}/d_c}$ (b) for the three engine configurations and both stratification versus ${t/\tau_b}$.}
\label{fig:fig06}
\end{figure}

More quantitatively, the vortex descent $z-z_0$ and the lifetime of the vortex dipole are shown in Figure \ref{fig:fig06}. Here, $z_0$ denotes the altitude of the vortex dipole at $t=0\text{ s}$, with values $z_{0,34\%}=20.74$ m; $z_{0,60\%}=22.76$ m and $z_{0,80\%}=16.21$ m. The lifetime of the vortex dipole is evaluated by computing the ratio between the minimum separation distance between the vortices and the vortex core average diameter \cite{Bouhafid24}. When this ratio reaches a value of 1, it indicates that the vortices are sufficiently close to interact and be destroyed by the Crow instability.

In terms of vortex descent, it is first observed that the descent is nearly linear across all configurations up to the point of complete vortex decay. This linearity is particularly evident under moderate stratification, where the descent reaches approximately $6b_0$, a value consistent with findings from previous studies such as the refs. \cite{Misaka14, Picot15, Saulgeot23, Bouhafid24}. Under strong stratification, the descent is reduced to around $2b_0$, consistent with the values reported in the previously cited studies. This reduction is attributed to stronger buoyancy forces, which counteract the vertical motion of the vortices.

Across the configurations, subtle differences in vortex descent behavior emerge under moderate stratification. The 34\% configuration exhibits the deepest descent, followed by the 60\%, while the 80\% configuration maintains the highest altitude. This deviation in the latter can be attributed to the fusion of the two vortices, which slows the descent before the first characteristic time, but can also be influenced by the proximity of the jet and its buoyancy effects. In this regard, the 34\% and 60\% configurations exhibit descent curves that fall lower, likely due to the proximity of the jet, which enhances descent through momentum transfer to the vortices \cite{Shirgaonkar07}. These observations are in line with the results of Saulgeot et al. \cite{Saulgeot23} up to approximately $5\tau_b$. Under strong stratification, the differences between configurations remain broadly similar to those observed in the moderate case, with the key distinction being that the 80\% configuration exhibits a greater descent relative to the 60\% case than it does under moderate stratification. The strong stratification is expected to enhance the instability of the six-vortex system, leading to its more rapid breakdown. This stratification level is challenging to compare directly with the findings of Saulgeot et al. \cite{Saulgeot23}, as in the present study, the early onset of the Crow instability rapidly disrupts the vortex dipole.

In terms of the vortex dipole's lifetime, stratification, as previously discussed, tends to shorten it by bringing the vortices together more rapidly. Overall, for both stratification levels and all three engine positions, the vortex tubes come into contact at approximately the same normalized times: around $4\tau_b$ for moderate stratification and $2.5\tau_b$ for strong stratification, except for the 60\% configuration under strong stratification, where vortex interaction occurs earlier, at $2\tau_b$. Ultimately, in all cases, the vortices are fully destroyed at nearly the same normalized time $t/\tau_b$, approximately $7\tau_b$ under moderate stratification and around $4\tau_b$ under strong stratification. These values are consistent with those reported by the refs. \cite{Bouhafid24,Bouhafid25} for the CRM configuration. The increased proximity of the vortex tubes in the 34\% and 80\% configurations also contributes to the intensification of elliptical instabilities observed at $t/\tau_b=3$, as their growth rate is strongly dependent on the inter-vortex distance \cite{Nomura06}. This effect remains visible at $t/\tau_b=5$, when the Crow instability drives the merging of the vortices.

Ultimately, quantifying the specific effect of engine placement on vortex behavior is challenging, given the complex interplay of competing aerodynamic phenomena. Nevertheless, the presence and proximity of the jet introduce additional buoyancy into the wingtip vortices and play a more prominent role in initiating elliptical instabilities. These instabilities are further amplified by local turbulence and the closeness of the jet to the vortex core and may interact with the Crow instability, altering its development.

\section{Impact on Contrail Properties}
\label{sec:Mphi}
\subsection{Spatial Distribution of Induced Contrails}

\begin{figure}[H]
\centering
\includegraphics[width=0.9\linewidth]{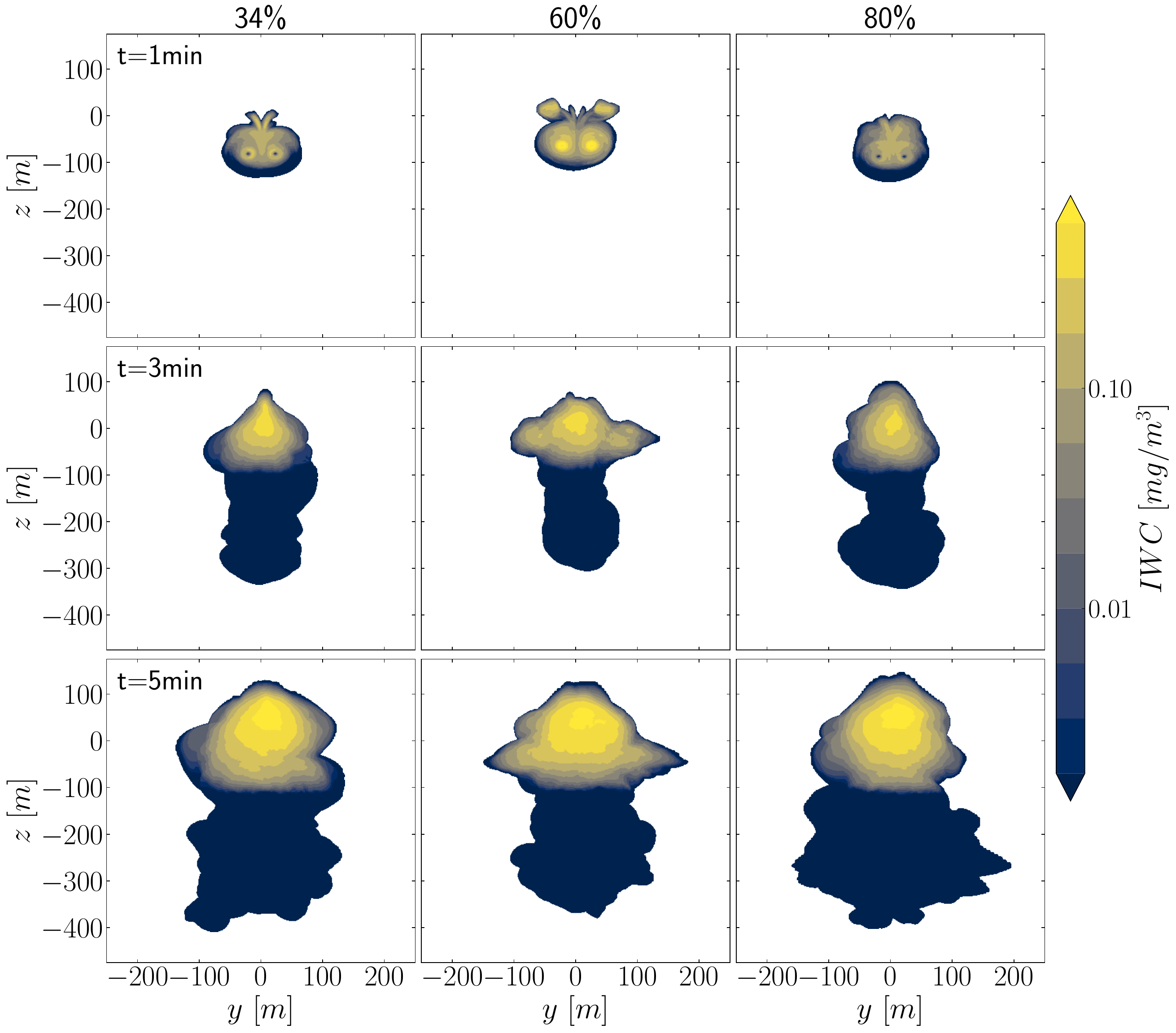}
\caption{Evolution of the ice plume from integrated IWC at 1, 3, 5 minutes for ${N_{BV}=0.012~\text{s}^{-1}}$ and the three engine positions.}
\label{fig:fig07}
\end{figure}

\begin{figure}[H]
\centering
\includegraphics[width=0.9\linewidth]{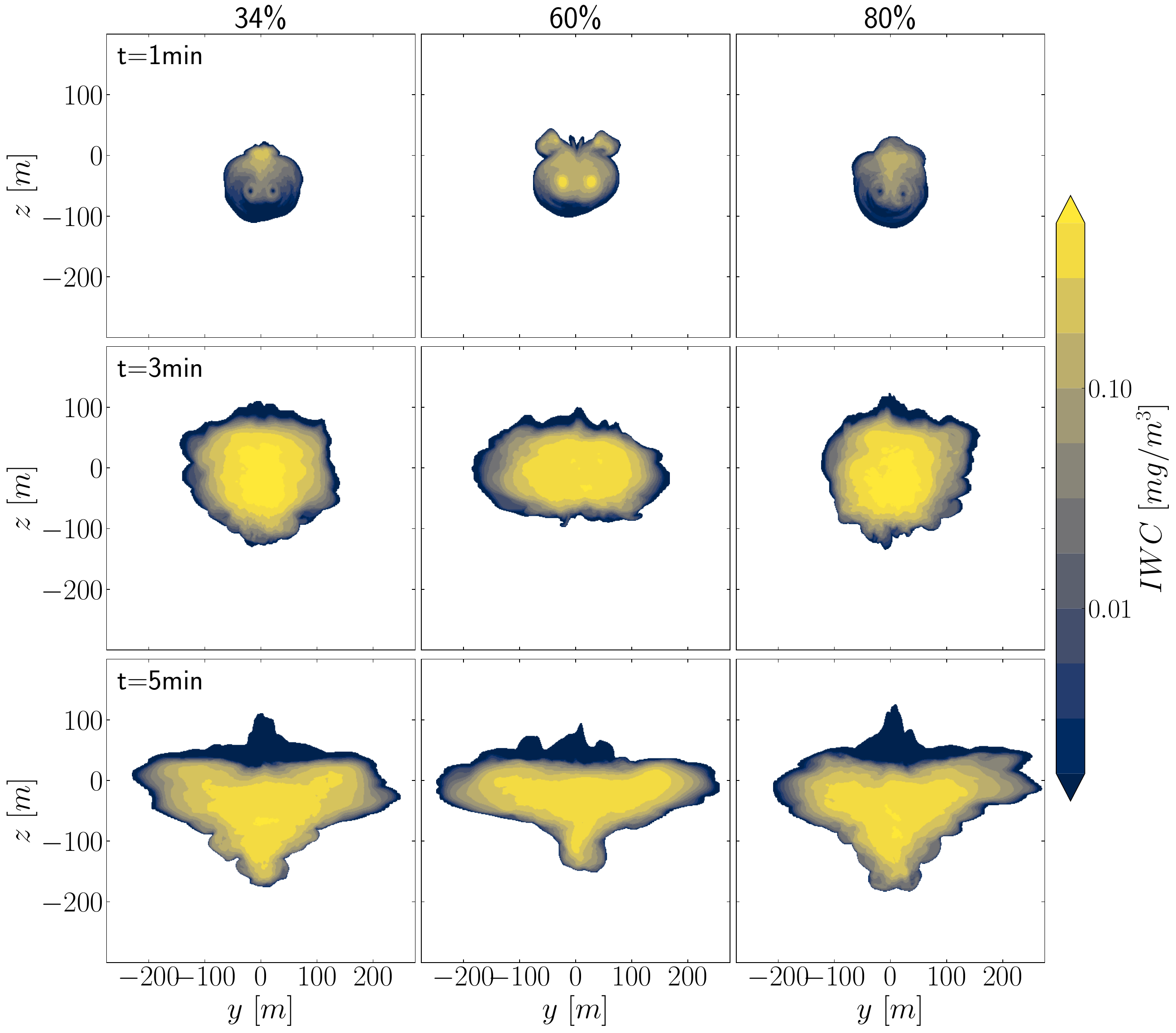}
\caption{Evolution of the ice plume from integrated IWC at 1, 3, 5 minutes for ${N_{BV}=0.030~\text{s}^{-1}}$ and the three engine positions.}
\label{fig:fig08}
\end{figure}

This section presents the microphysical results obtained for each contrail. The evolution of the ice plumes, under moderate and strong stratification, is respectively illustrated in Figures \ref{fig:fig07} and \ref{fig:fig08} using the integrated IWC field of the plume, as defined by Equation \ref{eq:eq15}. A masking is applied so that only the regions of the plume containing soot particles are displayed. The plume is shown at three moments in its lifetime: 1 minute, 3 minutes, and 5 minutes (rows) for each engine position (columns).

\begin{equation}
    IWC(y,z,t)=\frac{1}{L_x}\int \rho Y_{H_2O,s}(x,y,z,t)dx
    \label{eq:eq15}
\end{equation}

The evolution of the ice plume in each configuration is intrinsically linked to the aerodynamic behavior of the wingtip vortices. Across both stratification regimes, the 60\% engine configuration consistently displays the most distinct ice plume morphology compared to the 34\% and 80\% cases. This distinction arises primarily from the early-time interactions between the tail vortex and the vortex generated by flow separation (Figure \ref{fig:fig05}). These interactions retain a portion of ice particles near cruise altitude, leading to a more horizontally extended plume at later times and a less pronounced vertical spread. In contrast, the 34\% and 80\% configurations produce ice plumes that are more comparable, both being more vertically elongated and less laterally spread. 

Differences between moderate and strong stratification can be attributed to the distinct wake dynamics driven by baroclinic instabilities and buoyancy forces. For instance, at $N_{BV}=0.030\text{ s}^{-1}$ (Figure \ref{fig:fig08}), the vortices dissipate much more rapidly, causing the ice plume to remain at a higher altitude (Figure \ref{fig:fig06}). Consequently, the plume exhibits a broader horizontal spread and a reduced vertical extent compared to the moderate stratification case (Figure \ref{fig:fig07}). For the latter, the upper part of the plume, corresponding to the secondary wake, then contains almost all of the remaining ice. Most of the ice in the primary wake has sublimated during the compression associated with the descent of the vortex pair. The majority of very low IWC remaining values are likely attributable to numerical diffusion.

\begin{figure}[!ht]
\centering
\includegraphics[width=1.\linewidth]{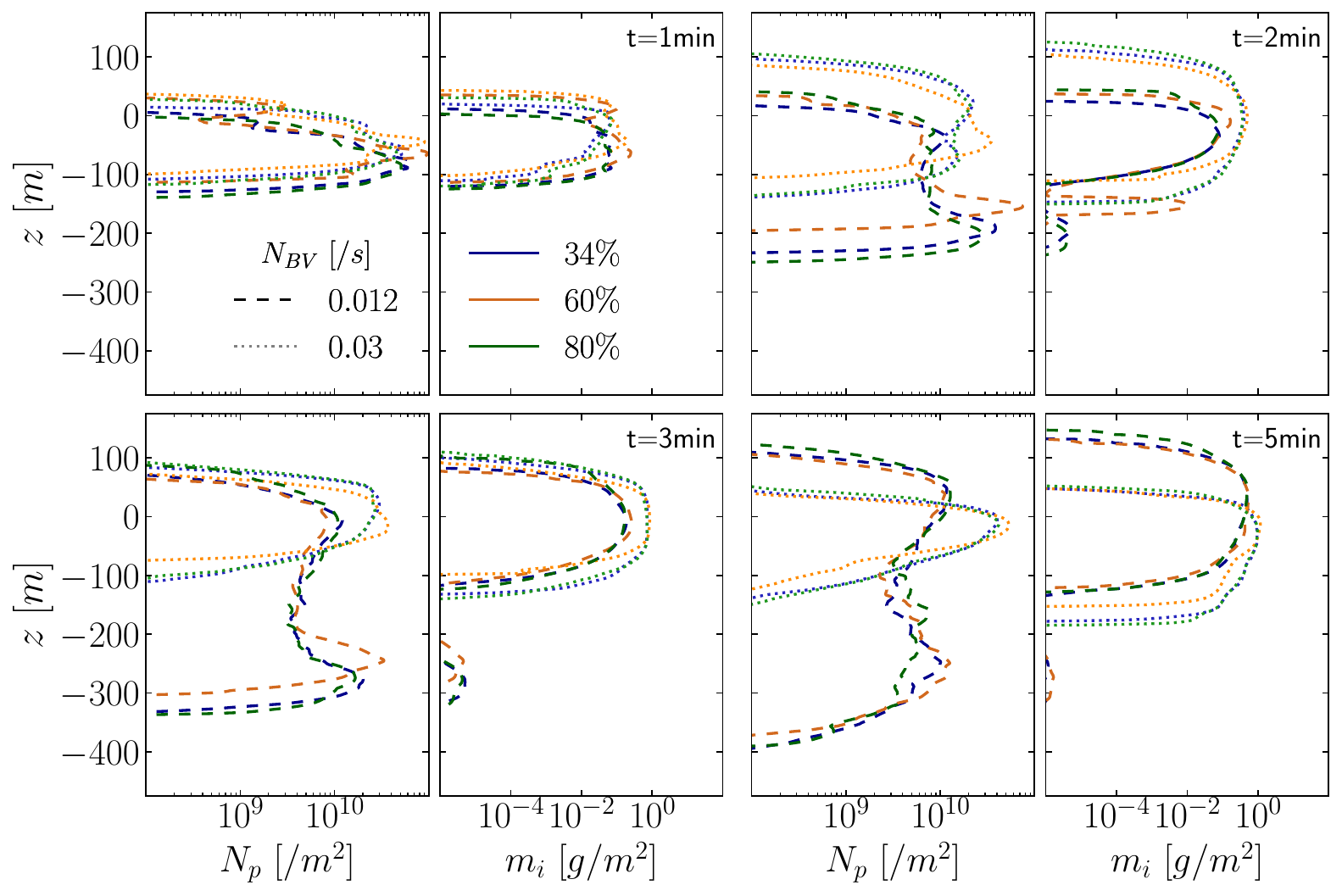}    
\caption{Vertical profiles of integrated soot density ${N_p(z,t)}$ (total particles, left) and ice mass ${m_i(z,t)}$ (right) for the three engine spacings and both stratification.}
\label{fig:fig09}
\end{figure}

To better analyze and quantify the impact of engine position on plume dimensions and dispersion, the vertical distributions of integrated total particle density $N_p(z,t)$ and ice mass $m_i(z,t)$ are examined in Figures \ref{fig:fig09}. These quantities are defined as follows:

\begin{equation}
    N_p(z,t)=\frac{1}{L_x}\iint N_p(x,y,z,t)dxdy
\end{equation}

\begin{equation}
    m_i(z,t)=\frac{1}{L_x}\iint \rho Y_{H_2O,s}dxdy.
    \label{eq:eq17}
\end{equation}

Slight differences are observed across the configurations, primarily driven by variations in particle distributions. An analysis of the vertical distribution of particles $N_p(z,t)$ shows that the 34\% and 80\% plumes are more spatially extended than the 60\% case for both stratification levels. This variation in spatial extent translates into differences in particle concentration, which are influenced by the number of particles entrained by the vortex dipole. 

The initial particle distribution plays a key role in shaping this entrainment. As shown in ref. \cite{Annunziata25}, the engine position significantly affects the initial spread of particles, with a greater or lesser proportion of soot entering the vortex region depending on the configuration. This effect is particularly pronounced for the 60\% case, where many particles are initially entrained into the vortex core. The consequences of this initial disparity are still apparent after 5 minutes of simulation in the lower part of the plume. For moderate stratification, the 60\% configuration exhibits a higher particle concentration peak in this area. In contrast, for the 34\% and 80\% cases, a more similar number of particles are entrained into the vortex dipole, consistent with the closer initial distributions reported in ref. \cite{Annunziata25}. However, both their final particle distributions differ due to differences in vortex dipole lifetime and the onset and strength of the Crow instability, which influences how the particles are dispersed within the lower part of the plume. These differences induced by the initial distribution are also evident in ref. \cite{Unterstrasser14}, where Gaussian and uniform jet initializations are compared.

These effects, however, have little impact on the ice mass distribution at $t=5~\text{min}$. For each stratification case, the distributions for the 34\% and 80\% configurations remain fairly similar, consistent with the similarities observed in particle distribution between these cases, particularly in the upper part of the plume. Under moderate stratification, the 60\% configuration exhibits an ice mass distribution that is broadly comparable to the other two, but with some notable differences. These include the emergence of a secondary peak in the distribution and a change in the amplitude of the main peak, features that are observed for both stratification levels. This behavior is primarily attributed to a different particle distribution than in the 34\% and 80\% configurations, leading to an ice plume that is more confined and less diluted, but more concentrated in ice content.

\begin{figure}[!ht]
\centering
\includegraphics[width=1.\linewidth]{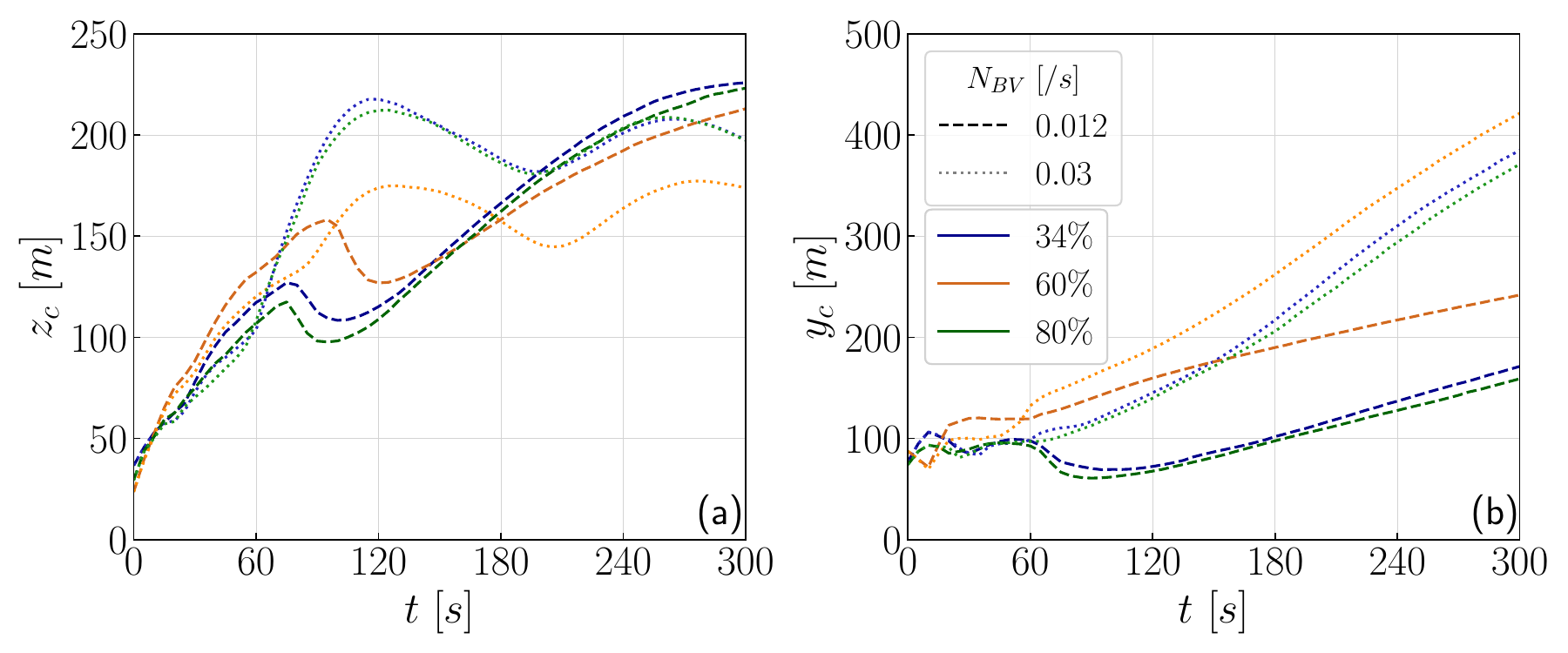}     
\caption{Contrail height ${z_c}$ (a) and width ${y_c}$ (b) for the three engine spacings, under both stratification levels, versus time ${t}$.}
\label{fig:fig10}
\end{figure}

Examining the morphology and distribution of the ice plumes suggests that engine position primarily influences their dimensions by altering particle dispersion. The characteristic sizes of each contrail, the height $z_c$ and width $y_c$, are then presented in Figure \ref{fig:fig10}. These two quantities are defined as the vertical and horizontal intervals containing 99.99\% of the total contrail ice mass. The sensitivity of the plume dimensions to the choice of criterion has been assessed using several definitions. The selected criterion exhibits low sensitivity.

As qualitatively observed earlier, the 34\% and 80\% configurations exhibit similar trends, while the 60\% configuration stands out, under both stratification values. In the vertical direction, the 34\% and 80\% contrails extend further than the 60\% case. This enhanced vertical spread is associated with the greater vortex descent as seen in Figure \ref{fig:fig06} for those configurations. Using the 34\% configuration as a reference, the relative differences, at $t=300$ s, are 5\% and 2\% for moderate stratification, and 12\% and 1\% for strong stratification, for the 60\% and 80\% configurations, respectively. 

In the horizontal direction, the relative differences are 40\% and 10\% for the 60\% and 80\% cases under moderate stratification. For strong stratification, they are 7\% and 4\%, respectively. As shown in Figures \ref{fig:fig07} and \ref{fig:fig08}, the presence of additional vortex structures in the 60\% configuration causes a portion of the ice plume to remain near cruise altitude, leading to a broader lateral spread. In contrast, for the other configurations, the degree of lateral dispersion varies depending on the proximity between the jet and the vortex, which influences the subsequent spreading. In terms of order of magnitude, particularly regarding contrail width, the results are in good agreement with values reported in the literature for B-777 \cite{UnterstrasserGorsch14}, consistent with the present study's configuration. Thus, the values are in good agreement with the magnitude reported in measurements \cite{Jessberger13,Schumann17}.

From a spatial perspective, modifying the engine position affects the spatial distribution of the induced contrails, influencing both their overall dimensions and particle concentration patterns. The spatial dispersion of particles is also influenced by the initial particle distribution, which determines the fraction of particles entrained by the vortex dipole. Despite these differences, the ice distribution exhibits a clear tendency toward uniformity across the engine configurations. This convergence is largely attributed to the similar overall morphology of the plumes, which drives the contrails toward comparable microphysical characteristics over time.

\subsection{Contrail Ice Content}

\begin{figure}[!ht]
\centering
\includegraphics[width=1.\linewidth]{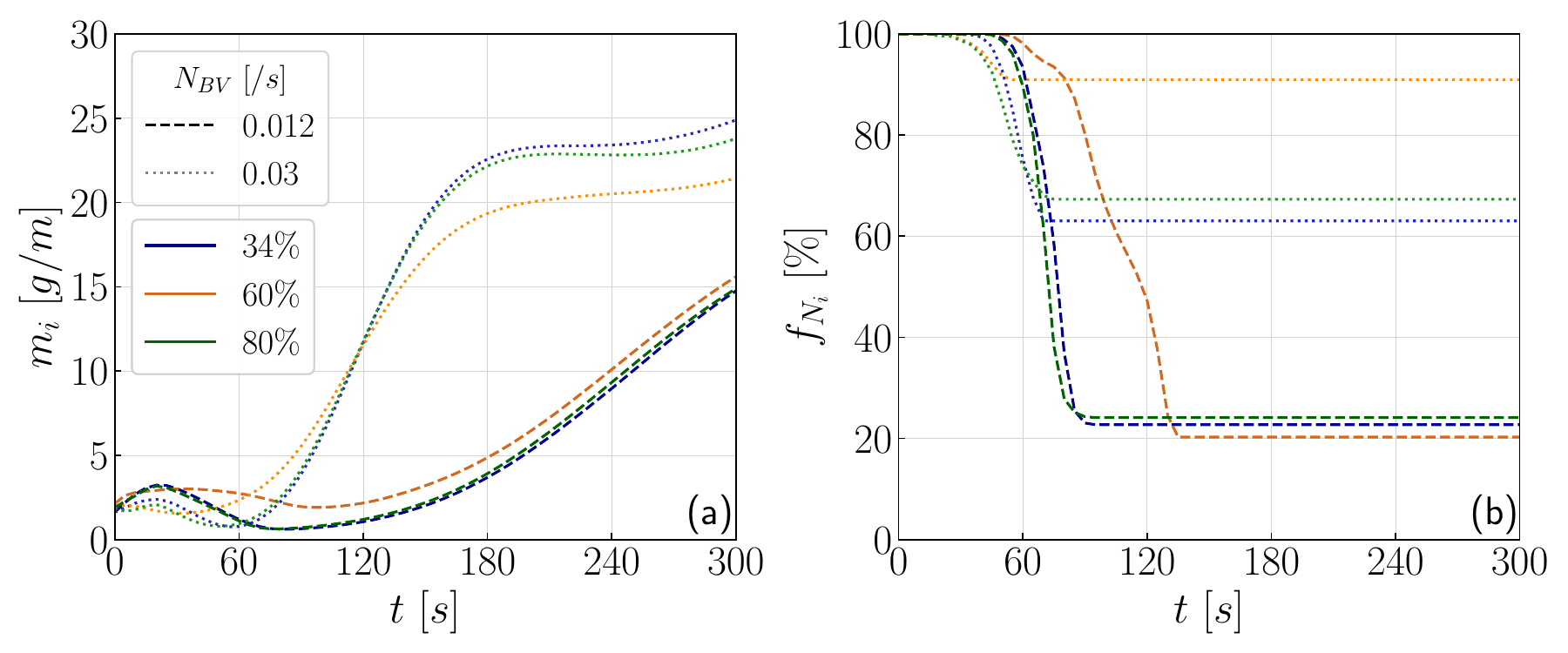}     
\caption{Total ice mass per flight meter ${m_i}$ (a) and total fraction of ice crystals ${f_{N,i}}$ (b), for the three engine spacings, and both stratification versus time ${t}$.}
\label{fig:fig11}
\end{figure}

As shown in the previous section, despite differences in vertical particle distribution and plume dimensions, the ice plumes exhibit relatively similar ice mass distributions across the different engine configurations. It is therefore of interest to examine the integrated ice properties of each induced contrail to assess whether this tendency toward uniformity persists over time. To this end, the total ice mass per flight meter, $m_i$, is presented in Figure \ref{fig:fig11}. It is obtained by $z$-integrating Equation \ref{eq:eq17}. 

Moreover, it is also of interest to examine the surviving crystals in each configuration, as this quantity is the most important for the contrail-to-cirrus transition \cite{Unterstrasser10pt1}. The fraction of surviving ice crystals, $f_{N_i}$, is then, jointly with the total ice mass, shown in Figure \ref{fig:fig11}. In the bulk framework adopted here, the transported particles $N_p$ are considered to be soot, unlike in the bulk models of \cite{Lewellen01,Unterstrasser10pt1}, where the transported particle density is assumed to correspond to the ice crystal density. The total soot number, here, is then conserved throughout the simulation, and no parameterization of crystal loss is included, in contrast to \cite{Unterstrasser10pt1,UnterstrasserSolch10}. Moreover, bulk models are known to reproduce ice mass reasonably well but to overestimate the number of ice crystals when these are directly transported \cite{Huebsch06,Unterstrasser08,UnterstrasserSolch10}. With this in mind, ice crystals are here defined as soot particles located within the cumulative region containing 99.99\% of the total ice in the simulation. This criterion excludes areas where ice is only present due to numerical diffusion (Figures \ref{fig:fig07} and \ref{fig:fig08}). It should be noted that a criterion based on particle radius did not prove satisfactory, as it was excessively sensitive for precisely this reason. By contrast, the mass-based criterion is reasonably representative of most of the contrail and exhibits comparatively low sensitivity across the various tests performed. This criterion is therefore not intended to provide an exact estimate of the number of sublimated crystals, something the bulk formulation cannot resolve, but rather to yield a representative order of magnitude for comparing the different engine positions.

Total ice mass effectively confirms the microphysical convergence observed. The 34\% and 80\% configurations are closely aligned, with 1\% and 5\% of deviation for moderate and strong stratification. The 60\% configuration remains the most distinct case, with a different evolution driven by the presence of the additional vortices. Its order of magnitude is comparable to that of the other engine positions, with 5\% and 14\% the differences with the 34\% configuration. The differences then arise due to the different spatial configurations of the plumes (Figures \ref{fig:fig09} and \ref{fig:fig10}). In terms of ice mass increase, these two configurations also exhibit the largest growth relative to their initial values at LES initialization. Initially, it was the 60\% and 80\% engine positions, respectively, in increasing order, that showed the highest ice mass. However, the 34\% configuration displays a greater relative increase compared to its initial state, regardless of the atmospheric stratification. This behavior can be attributed to an increased supply of water vapor to the plume through initial dilution (Figure \ref{fig:fig09}).

A few differences are found between engine positions. For moderate stratification, the discrepancies are the smallest: at the end of the simulations, the fractions of surviving ice crystals are 22, 20, and 24\% for the three engine positions, respectively. For strong stratification, the corresponding fractions are 63, 90, and 67\%. Overall, the differences between configurations remain limited, with the notable exception of the 60\% case under strong stratification, for which the vortices are destroyed more rapidly (Figure \ref{fig:fig06}), allowing a larger fraction of crystals to survive. These values for 60\% also explain the small differences observed for ice mass (Figure \ref{fig:fig10}). It is also worth noting that, for the other strongly stratified cases, the differences appear slightly larger than for the moderately stratified ones, indicating that when vortex lifetime is reduced, changes in dispersion may have a somewhat greater impact on microphysics. Nevertheless, the differences between configurations remain relatively small, suggesting that initial dilution plays a secondary role in determining the number of surviving crystals compared with vortex descent and atmospheric stratification.

Concerning the number of surviving crystals with the criterion used here, the orders of magnitude obtained in this study are broadly consistent with those reported in the literature for comparable ambient and aircraft conditions \cite{Lewellen01,Naiman11,UnterstrasserSolch10,Lottermoser25}. Furthermore, following ref. \cite{Lottermoser25} and computing the $z_\Delta$ value defined in their Equation 9, we obtain, for the three configurations, values of -13, -2, and -10 for moderate stratification, and 82, 96, and 83 for strong stratification. The values are broadly comparable with the cloud points shown in their Figure 9 for 110\%, as well as with the values reported in their study to compare with the literature \cite{Lewellen14pt1,Picot15}.

\begin{figure}[!ht]
\centering
\includegraphics[width=1.\linewidth]{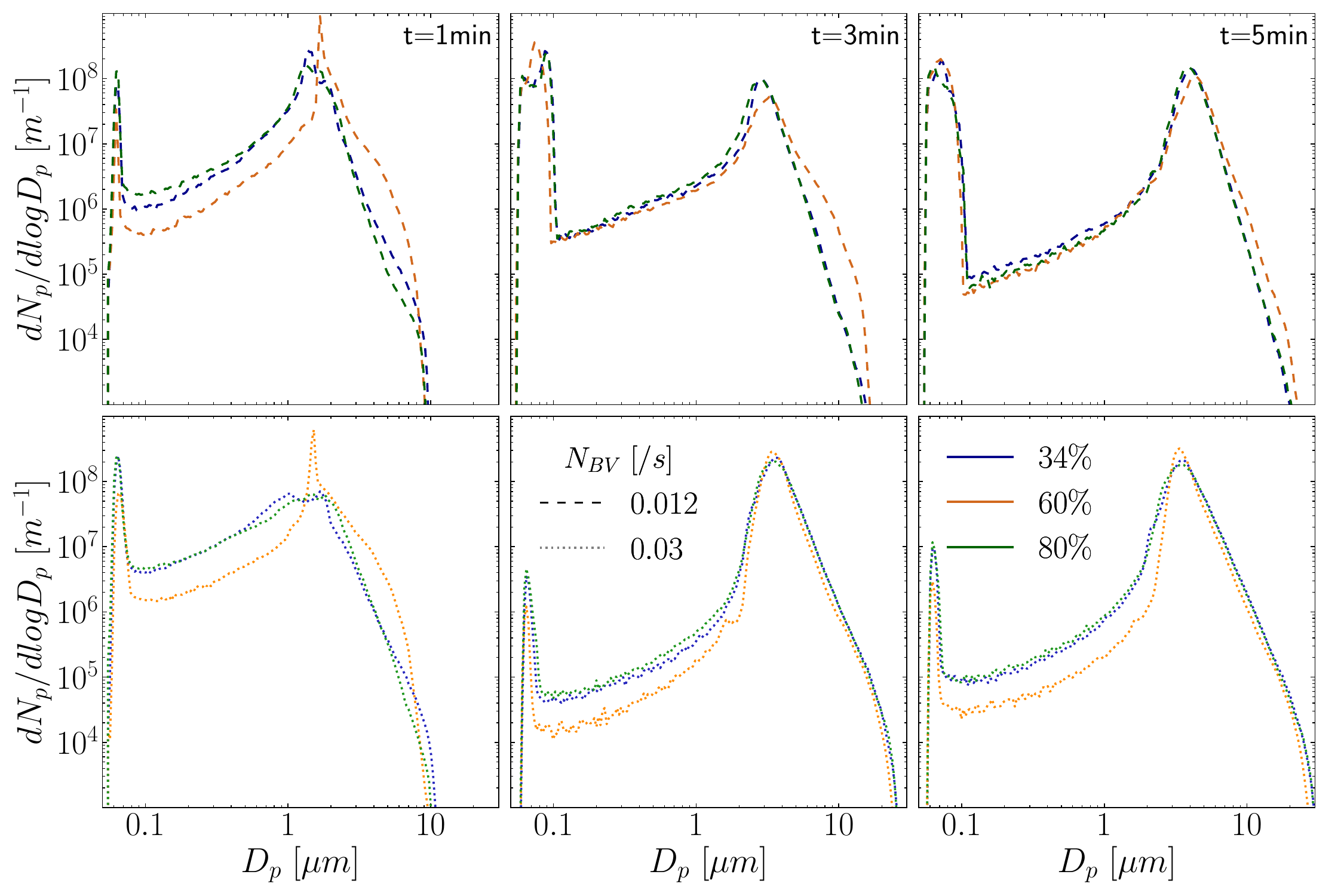}
\caption{Integrated total particle mean size distributions for three engine spacings under moderate (top) and strong (bottom) stratification, at ${t=1}$, 3, 5 min.}
\label{fig:fig12}
\end{figure}

To better understand the mechanism under this convergence, the integrated particle mean size class distributions are plotted as a function of particle diameter in Figure \ref{fig:fig12}. The top panels correspond to the moderate stratification case, while the bottom panels show results for strong stratification. It should be noted that the monodisperse assumption does not permit a fully accurate representation of the particle size distribution. The particle fields are therefore shown primarily for illustrative purposes, in order to provide physical insight into the mechanisms driving convergence among the different configurations.

The particle diameter distributions provide a more detailed characterization of the growth and ice crystal loss mechanisms, particularly through their distinctive shapes. Typically, the distributions exhibit a bimodal structure \cite{Schroder00}: a first peak at small diameters, corresponding to the sublimating ice crystals with radius close to that of dry soot, and a second peak at larger diameters, representing ice crystals in the secondary plume that have either risen or remained near cruise altitude, where sustained supersaturation favors continued growth. 

The key differences among the different engine positions lie in the relative amplitudes of these two peaks. In the early stages ($t=1$ min), the distributions are shaped by the initial entrainment of the plume by the wake vortices. In the 34\% and 80\% configurations, the initial ice crystal concentration is primarily governed by the dynamics of jet entrainment into the vortex system. This entrainment occurs more rapidly in the 80\% configuration, due to both the closer proximity of the jet to the vortex core and the faster descent of the vortex dipole (Figure \ref{fig:fig06}). These factors reduce the residence time of ice crystals in ice-supersaturated regions, limiting their growth. Consequently, fewer large-diameter crystals are formed, and those that do form undergo slightly more sublimation compared to the 34\% case. Conversely, the 60\% configuration shows the highest concentration of large particles, which is consistent with the retention of part of the plume at cruise altitude due to vortex interactions. In these regions, supersaturation with respect to ice allows crystals to grow more extensively. As the plume evolves ($t=3$ min and $t=5$ min), these trends gradually shift. The distributions become increasingly similar across the three cases, as vortex dynamics progressively dominate ice increase and loss over the effects of the initial dilution. The 60\% configuration still stands out for both stratifications, owing to the presence of the additional vortex structures.

As a result, no significant differences in microphysical properties are observed between engine positions, for the configurations studied here, at the end of the simulation. The main variations arise at the beginning of the vortex regime, where the flow dynamics still reflect the differences associated with each engine placement. These variations are subsequently outweighed by the effects of vortex dynamics and atmospheric stratification.

\subsection{Sensitivity to Variations in Ambient Relative Humidity}

To strengthen the conclusions drawn in the previous section regarding the influence of engine position, the sensitivity of the results is further assessed under a different ambient relative humidity with respect to ice. Whereas the previous results were obtained for an ambient $RH_i$ of 110\%, additional simulations are presented here for the 34\% and 80\% configurations under an ambient $RH_i$ of 120\%. The comparison is restricted to these two engine positions, since the 60\% configuration is affected by vortical structures associated with flow separation.

\begin{figure}[!ht]
\centering
\includegraphics[width=1.\linewidth]{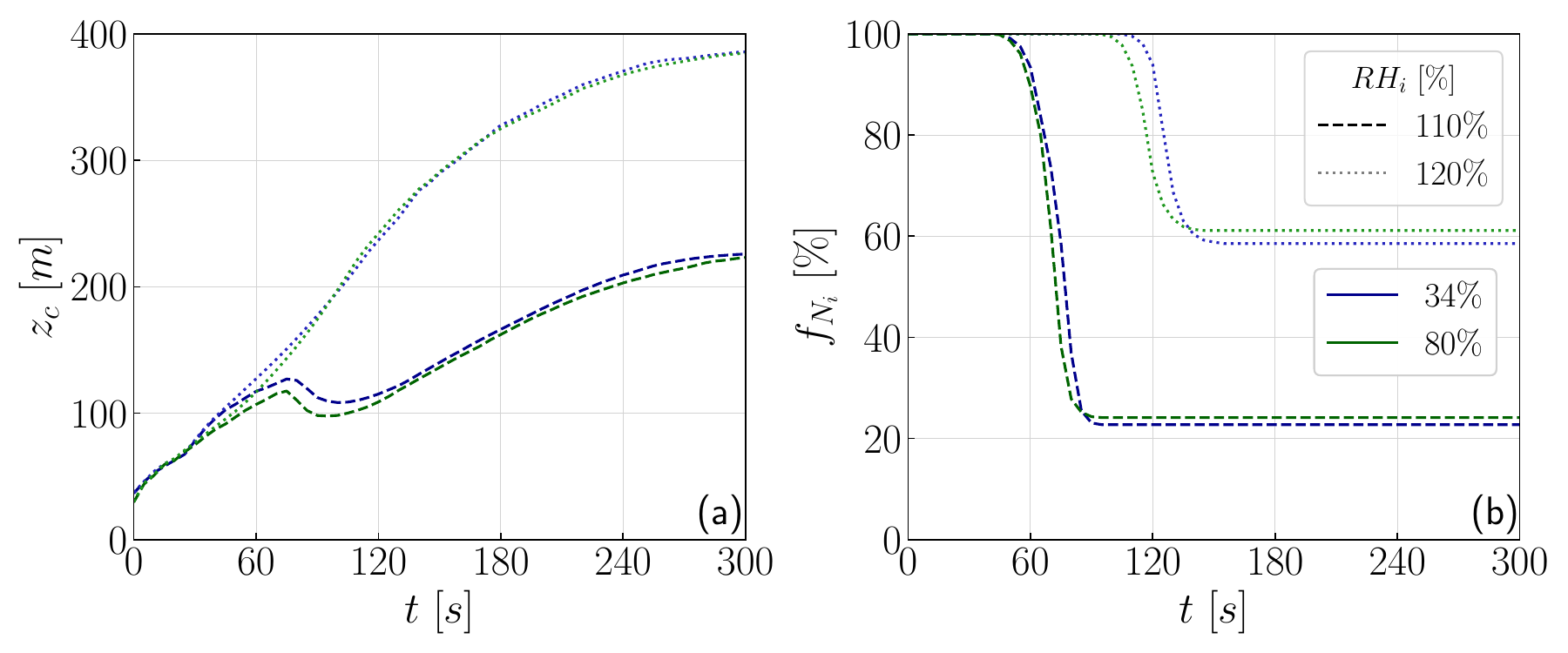}     
\caption{Contrail height ${z_c}$ (a) and total fraction of ice crystals ${f_{N,i}}$ (b) under moderate stratification, for ${RH_i=110\%}$ and ${120\%}$, versus time ${t}$.}
\label{fig:fig13}
\end{figure}

This additional comparison is particularly relevant because ambient relative humidity is a first-order parameter governing the early evolution of a contrail \cite{Unterstrasser10pt1} and, as shown in Figure \ref{fig:fig06}, one of the primary effects of engine position during the vortex regime is to modify both the descent and the lifetime of the vortex dipole. A change in ambient $RH_i$ may therefore have a significant impact, as a larger fraction of the contrail remains in supersaturated conditions, allowing more ice crystals to survive during vortex descent. Consequently, the differences in vortex descent induced by engine position may in turn affect the resulting contrail properties. To examine this, the vertical dimensions of the ice plume $z_c$, together with the fraction of surviving crystals $f_{N_i}$, are shown in Figure \ref{fig:fig13} for both $RH_i$.

The results indicate that increasing ambient relative humidity does not amplify the differences between engine positions, whether in terms of contrail height or the fraction of surviving ice crystals. Although not shown here, the same trend holds for other quantities previously examined, such as the plume width and the total ice mass. The relative differences obtained between the 34\% and 80\% configurations at $RH_i=120\%$ are on the order of 0.3\% for the plume height, against 2\% for $RH_i=110\%$. For the fraction of surviving ice crystals, values of 58\% and 61\% are obtained. The resulting discrepancy is fully consistent with that found at 110\%, remaining on the order of 2-3\%. These values, both in terms of plume dimensions and crystal survival, are consistent with those reported in \cite{Unterstrasser14,UnterstrasserGorsch14} for a B777, as well as with the results of \cite{Lottermoser25}. This agreement further supports the validity of the criterion used to define the ice plume.

Finally, variations in engine position exhibit relatively weak sensitivity to changes in ambient relative humidity. This finding not only reinforces the conclusions drawn in the previous section, particularly the tendency toward plume homogenization, but also highlights that, alongside atmospheric stratification and vortex dynamics, initial dilution effects (i.e., engine position) are a second-order parameter relative to ambient relative humidity. This contrasts with the jet regime, where both effects are of comparable magnitude \cite{Annunziata25}.

\section{Impact of Turbulence Randomization}
\label{sec:turb}

In this study, turbulent fluctuations rely on random numbers \cite{Castro13,Bouhafid24}. Between configurations, the jet turbulence corresponds to different realizations of the random variable, whereas the atmospheric turbulence is kept identical to facilitate comparison. Since it is the atmospheric turbulence that triggers the instabilities, it is interesting to examine another realization in order to assess how sensitive the results are to this random process and to compare this sensitivity with that associated with engine position.

\begin{figure}[!ht]
\centering
\includegraphics[width=1.\linewidth]{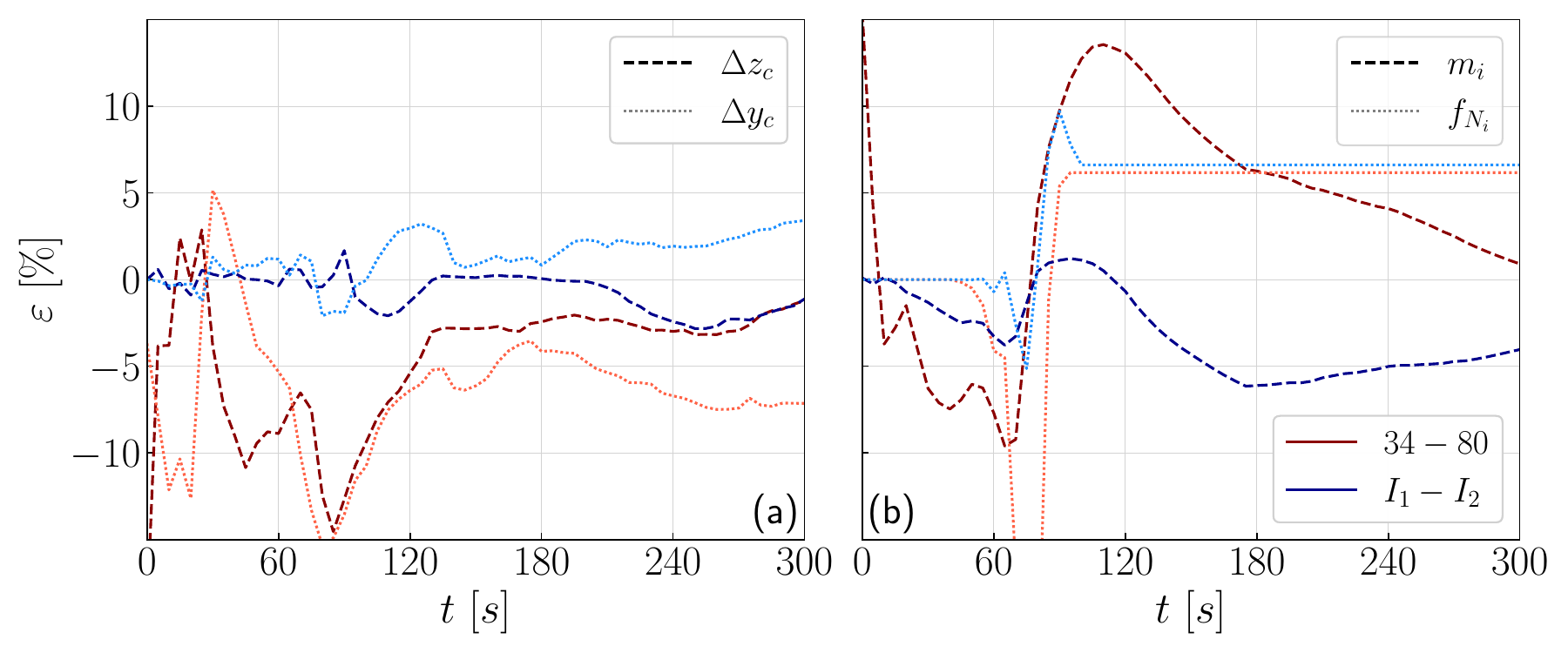}     
\caption{Relative deviation ${\varepsilon}$ between two fluctuation initializations for the 34\% configuration (blue) and between the old initializations for engine position (red), under moderate stratification, versus time ${t}$.}
\label{fig:fig14}
\end{figure}

To do so, an additional simulation is carried out for the 34\% configuration under moderate stratification, and $RH_i=110\%$. In this run, the jet and atmospheric fluctuations are regenerated to obtain a new realization of the random variables, while all other simulation parameters are kept identical to those of the previous simulation. To compare the two simulations, the relative deviations in contrail height and width (panel (a)), total ice mass, and the fraction of surviving ice crystals (panel (b)) are shown in Figure \ref{fig:fig14}. For comparison with the variations induced by engine position, the relative differences between the 34\% and 80\% configurations for $RH_i=110\%$ are also shown in the figure.

Regarding the relative differences associated with the new initialization (in blue), the largest relative deviations during the simulation are found in the ice mass and in the fraction of surviving ice crystals. They reach maximum values of -6\% and 10\%, respectively, and -4\% and -6.5\% at the end of the simulation. For the fraction of surviving crystals, the maximum deviation occurs during the sublimation phase, which therefore takes place at slightly different times in the two initializations. The differences in the horizontal and vertical extents of the two contrails are smaller, of the order of 1.5\% and 3\%, respectively. Interestingly, the ref. \cite{Unterstrasser10pt1} also reports differences associated with different fluctuation fields. They observed a 15\% difference, however, the study has been conducted in 2D.

These differences arise from the different fluctuation initializations. They modify the triggering of short-wavelength instabilities and also affect the onset of Crow instability. As a result, vortex decay is slightly altered, in particular, its timing. This, in turn, affects wake dispersion and microphysical properties. These differences are particularly interesting. Indeed, when compared to those typically associated with variations in engine position (in red), they show that the deviations previously observed between engine positions are of the same order of magnitude as those obtained by simply changing the turbulent fluctuation initialization. Since the orders of magnitude are comparable, this supports the conclusions of the previous section: the initial dilution differences induced in the jet regime are largely erased by larger-scale parameter effects.

\section{Conclusions}
\label{sec:ccl}

\subsection{Discussion and Outlooks}

Finally, it is important to acknowledge the limitations of this study, which are inherently linked to the numerical approach and modeling assumptions, as well as to highlight potential future developments. The strength of this study lies in its fully dynamic approach, which allows plume dilution processes to be captured consistently in both the near and far fields, for each engine position studied. The main limitations are primarily in the modeling of microphysics.

The microphysical model used in this study is a bulk scheme, which is known in the literature to exhibit certain limitations. As shown by refs. \cite{Huebsch06,UnterstrasserSolch10}, this approach reproduces the ice mass of a contrail reasonably well, but strongly overestimates the number of ice crystals surviving at the end of the vortex and dissipation regimes compared with Lagrangian or Eulerian sectional models. This discrepancy is even more pronounced when particles are assumed to be monodisperse within each mesh cell, as in this study. In the same way, ref. \cite{Gierens09} also shows that, within a bulk framework, it is challenging to introduce a source term capable of correcting this effect. In these studies, it is the ice crystals that are directly transported, and the term sublimation is applied to their concentration. In the present work, soot particles are transported, and they remain passive in the flow. Consequently, a mass-based filter must be used during post-processing to distinguish ice-bearing soot from ice-free soot.

This should be kept in mind when interpreting Figure \ref{fig:fig11}, which yields orders of magnitude comparable to those reported in the literature. However, the value may be biased, as it does not account for the heterogeneity of the contrail. The choice of a bulk model is ultimately justified by the need to strike a compromise between a full resolution of plume dilution and the computational cost associated with the microphysical model, particularly in the near-field. The use of a bulk monodisperse scheme also affects the near-field results, most notably by converting all particles into ice crystals. A more comprehensive microphysical treatment, particularly one accounting for polydispersity, would therefore be valuable. In light of the effects of ambient relative humidity and temperature identified in the near field, it is conceivable that changes in engine position may influence the number of ice crystals formed initially and, consequently, the number of crystals surviving farther downstream.

From a flow modeling perspective, some limitations can also be identified. The sensitivity of the Smagorinsky model constant $C_s$ has not been evaluated in the present work. In practice, selecting $C_s$ values that are too large or too small leads, respectively, to an over- or underestimation of the subgrid-scale eddy viscosity, and thereby to a more or less diffusive solution for a given grid resolution. Moreover, this may result in excessive eddy viscosity production within the vortex core. However, the results of ref. \cite{Bouhafid24} indicates that, with the value used here, jet-vortex interaction experiments are reproduced quite faithfully, giving us confidence in this choice. A possible refinement would be to employ a dynamic procedure that recomputes during the simulation \cite{Naiman11}. A possible limitation is the use of a second-order scheme, chosen to reduce computational cost. This may artificially diffuse the vortices and homogenize the mixture too rapidly. Future work could explore higher-order schemes, as in ref. \cite{Picot15}.

A further point worth discussing, closely related to the limitations of the RANS framework, concerns the extent to which such calculations truly require RANS-based initialization. In the present study, this initialization is important, notably to account for the dilution changes induced by variations in engine position, but also to properly represent the empennage vortices, which play a significant role in plume development \cite{Fabre02,Bouhafid25}. It would nevertheless be valuable to assess how the results would differ if the calculations had been initialized instead from a jet-vortex interaction LES \cite{Paoli13}, a hybrid RANS-LES approach \cite{Kolomenskiy18}, or an analytical model. More broadly, it remains to be determined to what extent the dynamical framework of the near field affects the subsequent microphysical evolution and, ultimately, the downstream contrail properties.

Lastly, another limitation of the present study concerns the vortex structures generated by flow separation on the CRM wing as a result of engine nacelle displacement. While these vortices exert minimal influence in the near-field, they have a more substantial impact in the far-field, particularly for the 60\% configuration, where they retain part of the plume near cruise altitude and consequently alter its evolution. This introduces a configuration-specific bias, whereas, in the 80\% case, the rapid merging of the vortices limits such effects.

\subsection{General Conclusion}

In the present study, Large-Eddy Simulations have been conducted over five minutes of contrail evolution, encompassing both the vortex and dissipation regimes. These simulations have been performed for three distinct engine positions, under two atmospheric stratification conditions (one moderate and one strong) and two ambient relative humidity, using an Eulerian microphysical bulk model. The LES computations have been initialized using RANS simulations carried out in the near-field of a realistic aircraft configuration, as presented in ref. \cite{Annunziata25}, incorporating the full 3D geometry (fuselage, wings, engines, and tailplane). The transition method of Bouhafid et al. \cite{Bouhafid24} has been employed to ensure continuity between the RANS and LES calculations. This approach enables consideration not only of the aircraft geometry but also of the complex jet-vortex interactions that arise immediately after ejection when the engine position is modified.

The results demonstrate that engine position, for the configurations and parameters employed, slightly influences the aerodynamic wake. The proximity of the jet to the wingtip vortex affects the development of vortex instabilities, notably by amplifying short-wavelength instabilities due to the jet's inherent turbulence. These amplified instabilities, in turn, alter the onset of the Crow instability, leading to variations in its timing and development across configurations. Additionally, the jet-vortex proximity impacts vortex dynamics more broadly: when the jet is closer to the vortex core, it contributes to a faster descent of the vortex pair and tends to reduce their separation, which further enhances the development of both short-wavelength instabilities and the Crow instability. However, the influence of engine position remains small. By contrast, stratification plays a much more important role.

Despite these differences, the three ice plumes tend toward uniformity across the stratification in terms of ice crystal number and ice content, with a larger ice mass increase in the cases that initially had lower values. The particle distribution is influenced primarily by the initial distribution, but also by the proximity of the particles to the vortex, which entrains the plume more or less rapidly. Even though these entrainment differences are noticeable in the early stages, the plumes tend to become similar in terms of size, ice mass, and number of surviving crystals. The differences observed are on the order of 5\% for each quantity for the 34\% and 80\% positions, which are the most comparable cases. The 60\% position is more specific owing to the additional vortices. The magnitude of the differences between configurations is therefore comparable to that obtained when changing the initialization of turbulent fluctuations. Moreover, changes in ambient relative humidity do not significantly affect the differences induced by engine position. This confirms that the initial dilution of the plume has only a minor influence compared with other key parameters such as vortex descent, ambient relative humidity, and stratification.

From a broader perspective, and in light of the conclusions drawn in ref. \cite{Annunziata25}, engine position, i.e, dilution, appears, for the parameters and configurations considered here, to exert a clear influence on near-field contrail properties such as the onset of formation, radius, and the distribution of ice crystals. By the end of the dissipation regime, however, contrail properties tend to become more uniform, and atmospheric conditions and vortex dynamics largely dominated the initial dilution differences. Nonetheless, the present results provide additional insight into contrail behavior and dispersion mechanisms. 

\section*{Acknowledgments}

The authors would like to thank Marc Massot for his thorough review and insightful suggestions, which significantly enhanced the clarity and overall quality of the manuscript. The authors gratefully acknowledge ONERA for funding this study.

\bibliographystyle{elsarticle-num} 
\bibliography{sample}

\end{document}